\documentclass[a4paper,11pt]{article}
\pdfoutput=1 

\usepackage{jheppub,braket} 

\usepackage[T1]{fontenc} 
\usepackage{longtable}
\usepackage{amsfonts,amsthm,amsmath,amssymb,graphicx,color}
\usepackage{xcolor,colortbl}
\usepackage{float}
\usepackage{graphicx}  
\usepackage{xcolor}
\usepackage{braket}
\usepackage{amssymb}   
\usepackage{amsmath,bm}
\usepackage{amsfonts}
\usepackage{bbm}

\usepackage{setspace}
\usepackage{mathrsfs}

\definecolor{Gray}{gray}{0.95}
\definecolor{LightCyan}{rgb}{0.88,1,1}

\newcommand{\be}{\begin{equation}}
\newcommand{\ee}{\end{equation}}
\newcommand{\bea}{\begin{eqnarray}}
\newcommand{\eea}{\end{eqnarray}}

\newcommand{\Tr}{\text{Tr}}

\newcommand{\ba}{\begin{eqnarray}}
\newcommand{\ea}{\end{eqnarray}}

\newcommand{\ca}{\mathcal}

\newcommand{\s}{\sqrt}

\newcommand{\ep}{\epsilon}
\newcommand{\prl}{Physical Review Letters}
\newcommand{\pra}{Physical Review A}
\newcommand{\prb}{Physical Review B}

 \def\ep{\epsilon}

\def\tr{\text{tr}}
\newcommand{\ex}[1]{\mathrm{e}^{#1}}

\newcommand{\pa}[1]{\left(#1 \right)}
\newcommand{\PA}[1]{\biggl(#1 \biggr)}

\newcommand{\br}[1]{\left[#1 \right]}
\newcommand{\BR}[1]{\Biggl[#1 \Biggr]}

\newcommand{\bb}[1]{\mathbb{#1}}

\newcommand{\abs}[1]{\left|#1\right|}

\newcommand{\ar}[1]{\xrightarrow[#1]{}}

\newcommand{\red}[1]{\color{red}{#1}}

\newcommand{\dg}[1]{#1^{\dagger}}
\newcommand{\fr}{\frac}

\def\be{\begin{equation}}
\def\ee{\end{equation}}
\def\ba{\begin{eqnarray}}
\def\ea{\end{eqnarray}}

 \def\ep{{\epsilon}}

 \def\a{{\alpha}}
 \def\ba{{\bar{\alpha}}}

 \def\D{{\Delta}}

 \def\b{{\beta}}
 \def\e{{\epsilon}}

\def\tr{{\text{tr}}}

\def\be{\begin{equation}}
\def\ee{\end{equation}}
\def\ba{\begin{eqnarray}}
\def\ea{\end{eqnarray}}

 \def\ep{{\epsilon}}

 \def\a{{\alpha}}
 \def\ba{{\bar{\alpha}}}

 \def\D{{\Delta}}

 \def\b{{\beta}}
 \def\e{{\epsilon}}

\def\tr{{\text{tr}}}

\def\be{\begin{equation}}
\def\ee{\end{equation}}
\def\ba{\begin{eqnarray}}
\def\ea{\end{eqnarray}}

 \def\ep{{\epsilon}}

 \def\a{{\alpha}}
 \def\ba{{\bar{\alpha}}}

 \def\D{{\Delta}}

 \def\b{{\beta}}
 \def\e{{\epsilon}}

\def\tr{{\text{tr}}}

\definecolor{Gray}{gray}{0.95}
\definecolor{LightCyan}{rgb}{0.88,1,1}

\def\Tr{{\text{Tr}}}
\newcolumntype{a}{>{\columncolor{Gray}}c}
\newcolumntype{b}{>{\columncolor{white}}c}

\title{\boldmath The quasi-particle picture and its breakdown after local quenches: mutual information, negativity, and reflected entropy}

\author[a]{Jonah Kudler-Flam,}
\author[b]{Yuya Kusuki,}
\author[a,c]{Shinsei Ryu}

\affiliation[a]{Kadanoff Center for Theoretical Physics, University of Chicago, Chicago, IL~60637, USA}
\affiliation[b]{Center for Gravitational Physics, Yukawa Institute for Theoretical Physics (YITP), Kyoto University, Kitashirakawa Oiwakecho, Sakyo-ku, Kyoto 606-8502, Japan.}
\affiliation[c]{James Franck Institute, University of Chicago, Chicago, Illinois 60637, USA}

\emailAdd{jkudlerflam@uchicago.edu}
\emailAdd{yuya.kusuki@yukawa.kyoto-u.ac.jp}
\emailAdd{ryuu@uchicago.edu}

\abstract{
We study the dynamics of (R\'enyi) mutual information, logarithmic negativity, and (R\'enyi) reflected entropy after exciting the ground state by a local operator. Together with recent results from Ref.~\cite{2020arXiv200105501K}, we are able to conjecture a close-knit structure between the three quantities that emerges in states excited above the vacuum, including both local and global quantum quenches. This structure intimately depends on the chaoticity of the theory i.e.~there exist distinct sets of equivalences for integrable and chaotic theories. For rational conformal field theories (RCFT), we find all quantities to compute the quantum dimension of the primary operator inserted. In contrast, we find the correlation measures to grow (logarithmically) without bound in all $c>1$ conformal field theories with a finite twist gap. In comparing the calculations in the two classes of theories, we are able to identify the dynamical mechanism for the breakdown of the quasi-particle picture in 2D conformal field theories. Intriguingly, we also find preliminary evidence that our general lessons apply to quantum systems considerably distinct from conformal field theories, such as integrable and chaotic spin chains, suggesting a universality of entanglement dynamics in non-equilibrium systems.
}

\begin{document} 
\maketitle
\flushbottom

\section{Introduction}

Integrable quantum mechanical systems have as many conserved quantities as degrees of freedom. This highly constrains the dynamics, leading to significant progress in their solutions. A subtle limit occurs when progressing to quantum field theories where one has an infinite number of degrees of freedom. Certainly, one needs an infinite number of conserved quantities for the theory to be integrable, but is infinity enough? This tension between infinities is particularly sharp in two-dimensional conformal field theory (CFT) where there is always an infinite number of conserved charges (the quantum KdV charges) associated to the infinite dimensional Virasoro symmetry \cite{sasaki1988, 1996CMaPh.177..381B}. Certain conformal field theories, for example the unitary minimal models, show features of integrability, while others, such as those possessing holographic duals, show features of chaoticity \cite{Roberts2015}. For massive quantum field theories, integrability may be defined by the factorization of the S-matrix into $2\rightarrow 2$ scattering i.e.~no particle production, but the S-matrix is not well-defined in conformal field theories, so it is not clear how the notions of integrability translate to 2D CFTs. In particular, how does the breakdown of integrability manifest itself? One edifying approach to this question is studying far-from-equilibrium dynamics, in particular, entanglement dynamics.

Though characterizing the flow and generation of entanglement in interacting non-equilibrium situations seems intimidating, remarkably, universal features have been shown to emerge. In particular, the \textit{quasi-particle picture} has been proposed to quantitatively capture the evolution of entanglement entropy in integrable quantum systems in the scaling limit \cite{2005JSMTE..04..010C,2017PNAS..114.7947A,2018ScPP....4...17A}. This phenomenological description posits that in highly excited states, entanglement is purely bipartite and carried by local quasi-particles pairs. When a quasi-particle is within a subregion $A$ and its partner is outside of $A$, then $A$ is entangled with its complement and has nontrivial entanglement entropy. 
Their velocities and entanglement content may be obtained using thermodynamic Bethe ansatz methods. While the quasi-particle picture provides an elegant universal description of integrable models, it fails to capture entanglement dynamics in more generic systems. In this paper, we demonstrate the mechanism for this breakdown in 2D CFT.

In this approach, we study mixed state correlation measures following the insertions of local operator upon the ground state. The three quantities of interest are defined as follows:
\begin{enumerate}
    \item The \textbf{R\'enyi mutual information} is a one-parameter family of correlation measures defined as
    \begin{align}
        I^{(n)}(A:B) \equiv S^{(n)}(A) + S^{(n)}(B) - S^{(n)}(A\cup B)
    \end{align}
    where $S^{(n)}(\Omega)$ is the R\'enyi entropy of the reduced density matrix $\rho_{\Omega}$
    \begin{align}
        S^{(n)}(\Omega) \equiv \frac{1}{1-n} \log \Tr \rho_{\Omega}^n.
    \end{align}
    The von Neumann limit corresponds to taking the R\'enyi index, $n$, to one.
    \item The \textbf{logarithmic negativity} is a proper measure of entanglement for mixed states defined as \cite{PhysRevLett.77.1413,1996PhLA..223....1H,1999JMOp...46..145E,2000PhRvL..84.2726S,2002PhRvA..65c2314V,2005PhRvL..95i0503P}
    \begin{align}
        \mathcal{E}(A:B) \equiv \log \left|\rho_{AB}^{T_B} \right|_1
    \end{align}
    where $\left|\cdot \right|_1$ is the trace norm and $\cdot^{T_B}$ is the partial transpose operation. 
    \item The \textbf{R\'enyi reflected entropy} is the R\'enyi entropy of a ``canonical purification'' of a generic density matrix \cite{2019arXiv190500577D}
    \begin{align}
        S_R^{(n)}(A:B) = S^{(n)}(AA^*),
    \end{align}
    where the purification is defined on a doubled Hilbert space $\mathcal{H}_A \otimes \mathcal{H}_{A^*}\otimes \mathcal{H}_B\otimes \mathcal{H}_{B^*}$ 
    \begin{align}
        \rho_{AB} = \sum_i p_i \ket{\psi_i}\bra{\psi_i}_{AB} \rightarrow \ket{\sqrt{\rho_{AB}}}_{AA^*BB^*} \equiv \sum_{i}\sqrt{p_i}\ket{\psi_i}_{AB} \ket{\psi_i^*}_{A^*B^*}.
    \end{align}
In practice, this purified state can be realized by the analytic continuation of the even integer $m$ to one for the following state,
\begin{equation}\label{eq:purification}
\ket{\rho_{AB}^{m/2}}_{AA^*BB^*} \equiv \sum_{i}p_i^{m/2}\ket{\psi_i}_{AB} \ket{\psi_i^*}_{A^*B^*}.
\end{equation}
\end{enumerate}

\paragraph{Emergent web of correlation measures}
In integrable systems, the logarithmic negativity was argued to equal half of the R\'enyi mutual information following a quasi-particle picture \cite{2018arXiv180909119A}
\begin{align}
    \Delta \mathcal{E}(A:B) = \frac{ \Delta I^{(1/2)}(A:B)}{2} ,
    \label{cal_al_prop}
\end{align}
where $\Delta$ means the change with respect to the ground state\footnote{While the $\Delta$'s were not explicitly written in Ref.~\cite{2018arXiv180909119A}, they were implied because all discussion was about states highly excited above the ground state.}. This was argued through the existence of infinitely-living quasi-particles; thus, a priori, it may break down for chaotic systems. While this is already a very nice unification of two seemingly different correlation measures, we propose that there are significantly richer connections. In particular, we propose that the reflected entropy also obeys the quasi-particle picture and is equivalent to the mutual information for all values of the replica index
\begin{align}
    \Delta S_R^{(n)}(A:B) = \Delta I^{(n)}(A:B).
    \label{sr_I_conj}
\end{align}
Furthermore, though (\ref{cal_al_prop}) is expected to break down in chaotic theories, we conjecture that an analogous relation may be made regardless of the theory as long as the system is sufficiently excited and in the scaling limit
\begin{align}
    \Delta \mathcal{E}(A:B) = \frac{ \Delta S_R^{(1/2)}(A:B)}{2}.
    \label{ln_sr_conj}
\end{align}
This has previously been shown to hold in 2D CFTs possessing holographic duals \cite{PhysRevLett.123.131603}, but it is unexpected and interesting that it holds for both rational and irrational theories with finite central charge. We note that it cannot hold for generic quantum states because the reflected entropy is sensitive to both quantum and classical correlations while the negativity is only sensitive to quantum correlations, so the universality only emerges in highly excited states.

There are several methods for probing non-trivial dynamics of these correlation measures in conformal field theory, all of which require the computation of $n>3$-point correlation functions\footnote{The exceptions are semi-infinite intervals after Calabrese-Cardy local and global quenches and local operator quenches as these only require two and three-point functions respectively, though these are not interesting because the total state is pure, so the negativity, by definition, is simply equal to the R\'enyi entropy at index $1/2$ and the reflected entropy, by definition, is just twice the von Neumann entropy.}. Thus, each one of these computations requires the dynamical input of the full operator content of the theory of interest. Generally, in the past, only the universal contributions have been evaluated. This only reproduces quasi-particle dynamics. In fact, for quenches prepared by a path integral on a manifold that may be conformally mapped to the upper half plane\footnote{These include, for example, inhomogeneous global quenches \cite{2008JSMTE..11..003S}, finite-size global quenches \cite{2014PhRvL.112v0401C,2016arXiv160407830M}, splitting local quenches \cite{2019JHEP...03..165S}, double local quenches \cite{2019arXiv190508265C}, and Floquet CFT \cite{2018PhRvB..97r4309W,2018arXiv180500031W}.}, the quantities for disjoint intervals are completely agnostic to the specific conformal map \cite{2015PhRvB..92g5109W, 2020arXiv200105501K}
\begin{align}
    \mathcal{E} = -\frac{c}{8} \log\left(\frac{\eta_{1,4}\eta_{2,3}}{\eta_{1,3}\eta_{2,4}} \right), \quad I^{(n)} = S_R^{(n)} = -\frac{c(n+1)}{12n} \log\left(\frac{\eta_{1,4}\eta_{2,3}}{\eta_{1,3}\eta_{2,4}} \right) ,
\end{align}
where $\eta_{i,j}$ are various conformally invariant cross-ratios whose definitions may be found in e.g.~Ref.~\cite{2020arXiv200105501K}.
Therefore, (\ref{cal_al_prop}) and (\ref{sr_I_conj}) will hold for any quantum quench in this class. While this begins to explain why these confluences are natural from the CFT perspective, it begs several questions: (i) What is so special about $n= 1/2$ in \eqref{cal_al_prop} when all R\'enyi's are proportional? (ii) What happens when we move beyond the universal contribution and account for theory dependence? (iii) What happens if we probe dynamics that cannot be computed by correlation functions of twist-fields on the upper half plane? We will provide definitive answers to all three of these questions.

\subsection{Summary of main results}
\paragraph{Quasi-particle picture for negativity in CFT}

In rational CFTs, we find that the negativity after the local quench is fully captured by the quasi-particle picture with quasi-particles moving at the speed of light. The entanglement content of the quasi-particle created by the local operator is the logarithm of the quantum dimension. Using this fact and prior results for R\'enyi mutual information, we are able to confirm (\ref{cal_al_prop}) for RCFT and show that analogous statements for other R\'enyi entropies would be inconsistent. Furthermore, by computing the R\'enyi reflected entropy, we are able to provide evidence for our conjecture (\ref{sr_I_conj}) and confirm (\ref{ln_sr_conj}) in RCFT.

\paragraph{Breakdown of quasi-particle picture}

When progressing to $c>1$ CFTs with finite twist gap\footnote{We sometimes refer to this class of CFTs as \textit{pure} because their complete symmetry algebra is $Vir \times \overline{Vir}$.}, we demonstrate the breakdown of the quasi-particle picture for all three quantities. Namely, in the Regge limit, the dominant operator exchange in the cross-channel conformal block is no longer the identity operator, as it was for RCFT.
This fact essentially destroys the notion of local propagating quasi-particles and leads to logarithmic growth in all correlation measures. In this way, we find (\ref{cal_al_prop}) and (\ref{sr_I_conj}) to break down, but (\ref{ln_sr_conj}) to persist. This mechanism for the breakdown of integrability is intimately tied to the fact that these CFTs have an infinite number of primary fields.

\paragraph{Role of backreaction for holographic negativity}

The holographic prescription for logarithmic negativity involves computing the area of a gravitating entanglement wedge cross-section \cite{2019PhRvD..99j6014K, PhysRevLett.123.131603}, a highly nontrivial gravitational task. However, to leading order, the backreaction may be accurately accounted for by simply computing the entanglement wedge cross-section without backreaction \cite{2020JHEP...01..031K}. Explicit checks of this approximation are few, and we provide the first check for dynamical nonsymmetric spacetimes. We find that while the leading approximation correctly predicts the logarithmic growth of negativity, the overall coefficient is corrected due to the gravitational interactions between a falling particle and the tensionful entanglement wedge cross-section. 

\subsection{Organization}
The rest of the paper is organized as follows. In Section \ref{review}, we review the construction of local operator quenches in conformal field theory and how to compute the correlation measures using various twist-field (replica trick) formalisms. In Section \ref{RCFT}, we study rational CFTs, deriving a quasi-particle picture. In Section \ref{pureCFT}, we study a $c>1$ CFTs with finite twist gap and demonstrate the mechanism for the breakdown of the quasi-particle picture. In Section \ref{backreaction_sec}, we compare our results to the holographic description of logarithmic negativity, characterizing the effect of backreaction in the bulk. In Section \ref{lattice}, we simulate integrable and chaotic lattice models, finding strong similarities to the CFT results. Some subtleties regarding conformal blocks are addressed in the Appendix.

\section{Review}
\label{review}

\begin{figure}
    \centering
  \includegraphics[width = .75\textwidth]{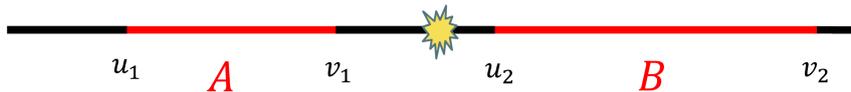}
 \caption{We study the setup $0< u_2<-v_1<-u_1<v_2$. We excite the vacuum by inserting a local operator at the origin at $t=0$.}
 \label{fig:setup}
\end{figure}

For the remainder of the paper, we focus on the following mixed state,
\begin{equation}\label{eq:state}
\rho_{AB}=\tr_{\overline{AB}} \ket{\Psi(t)} \bra{\Psi(t)},
\end{equation}
where $\ket{\Psi(t)}$ is a time-dependent pure state prepared by inserting a local Virasoro primary operator on the vacuum
\begin{align}
\label{state}
    \ket{\Psi(t)}=\s{\ca{N}}\ex{-\ep H-iHt} O(0)\ket{0},
\end{align}
where $\mathcal{N}$ is the normalization and $\epsilon$ is a UV regulator. $A$ and $B$ are two disjoint intervals as shown in Fig.~\ref{fig:setup}.

\subsection{Replica tricks}

\paragraph{Mutual Information}

\begin{figure}
    \centering
    \includegraphics[width = \textwidth]{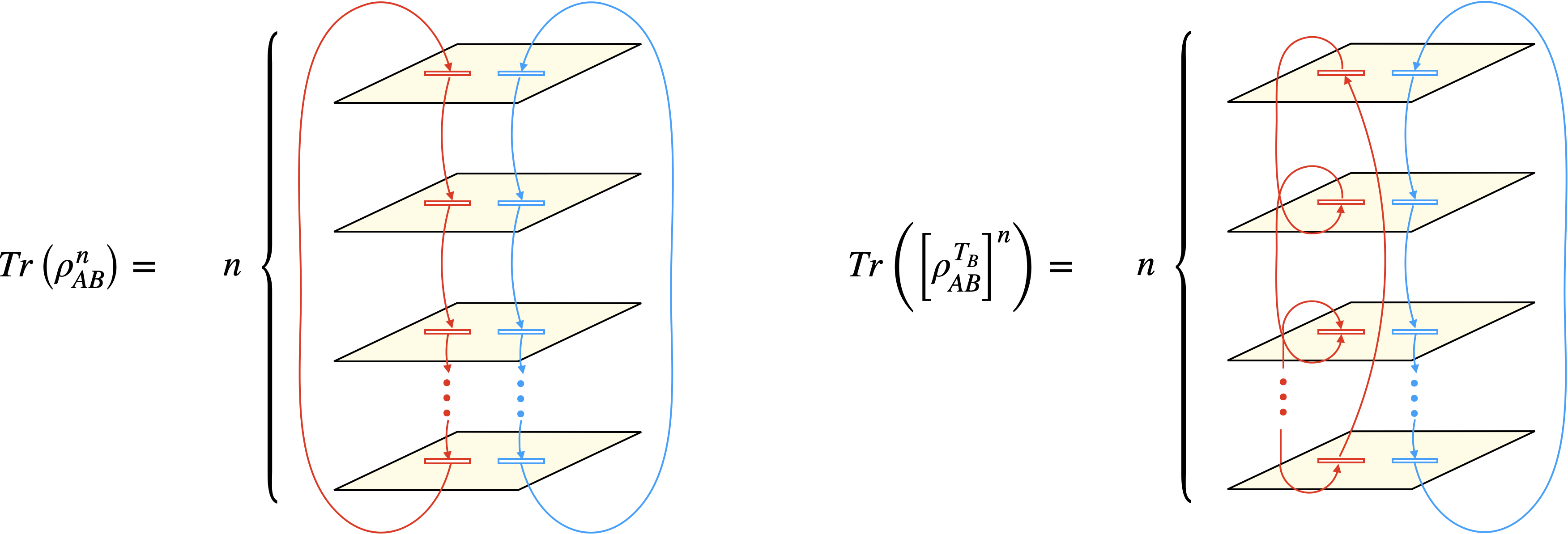}
    \caption{Left: the path integral representation for the moments of the reduced density matrix $\rho_{AB}$. The $n$ sheets are glued cyclically in the same order for both region $A$ (red) and region $B$ (blue). Right: the path integral representation for the moments of the partially transposed reduced density matrix. Note that region $A$ is now glued anti-cyclically. This is the action of the partial transpose.}
    \label{replica_LN_MI}
\end{figure}
We compute the mutual information using the replica trick in the path integral representation (Fig.~\ref{replica_LN_MI}).
In particular, the formulation of the entanglement entropy for a single interval after a local quench was given in Ref.~\cite{2014PhRvL.112k1602N}.
The generalization of their formulation to two disjoint intervals $A\cup B$ is straightforward. We use correlation functions of twist operators to compute the multi-sheeted partition function
\begin{equation}\label{eq:RenyiMI}
S^{(n)}(AB)[O]=
\fr{1}{1-n}\log \fr{
	   \Braket{\sigma_{n}(u_1)\sigma_{n}^{-1}(v_1)  O^{\otimes n}(w_1,\bar{w}_1)  \dg{{O^{\otimes n}}} (w_2,\bar{w}_2)   \sigma_{n}(u_2) \sigma_{n}^{-1}(v_2) }_{\text{CFT}^{\otimes n}}  }
{\PA{\Braket{ O(w_1,\bar{w}_1)  \dg{O} (w_2,\bar{w}_2)}}^n },
\end{equation}
where we abbreviate $V(z,\bar{z})\equiv V(z)$ if $z\in\bb{R}$ and the operators $O$ are inserted at
\begin{equation}
w_1=t+ i \e, \ \ \ 
\bar{w}_1=-t+ i \e, \ \ \ 
w_2=t- i \e, \ \ \ 
\bar{w}_2=-t- i \e. \ \ \ 
\end{equation}
We denote $O^{\otimes n}\equiv O\otimes O\otimes\cdots \otimes O$ as the operator on $n$ copies of the CFT ($\text{CFT}^{\otimes n}$). 
The twist operator $\sigma_n$, a byproduct of the $\mathbb{Z}_n$ orbifold, is a Virasoro primary with dimensions $h_{n}=\bar{h}_{n}=\fr{c}{24}\pa{n-\fr{1}{n}}$. The mutual information is then defined by
\begin{equation}\label{eq:MIdef}
I(AB)[O]=\lim_{n \to 1}\pa{S^{(n)}(A)[O]+S^{(n)}(B)[O]-S^{(n)}(AB)[O]},
\end{equation}
where $S(A)$ ($S(B)$) is the entanglement entropy for the subsystem $A$ ($B$).


\paragraph{Logarithmic Negativity}
The logarithmic negativity in conformal field theory was first considered in Ref.~\cite{2012PhRvL.109m0502C}.
In the path integral representation (Fig.~\ref{replica_LN_MI}), the corresponding correlation function of twist fields is given by exchanging the twist and anti-twist operators for one interval with respect to the correlation function for entanglement entropy (\ref{eq:RenyiMI})
\begin{equation}
\sigma_{n}(u_2) \sigma_{n}^{-1}(v_2)  \to  \sigma_{n}^{-1}(u_2) \sigma_{n}(v_2).
\end{equation}
It is this exchange that implements the partial transposition.
This means that the moments of the negativity for disjoint intervals following a local operator quench can be evaluated as
\begin{equation}\label{eq:RenyiN}
\ca{E}^{(n)}(A:B)[O]
=\log \fr{
	   \Braket{\sigma_{n}(u_1)\sigma_{n}^{-1}(v_1)  O^{\otimes n}(w_1,\bar{w}_1)  \dg{{O^{\otimes n}}} (w_2,\bar{w}_2)   \sigma_{n}^{-1}(u_2) \sigma_{n}(v_2) }_{\text{CFT}^{\otimes n}}  }
{\PA{\Braket{ O(w_1,\bar{w}_1)  \dg{O} (w_2,\bar{w}_2)}}^n },
\end{equation}
where the normalization is determined by $\Tr \rho_{AB}^{T_B} =1$. To take the trace norm, we must analytically continue {\it even} integers $n_e$ to one\footnote{See Ref.~\cite{Tamaoka2019} for an interesting discussion of the odd moments.},
\begin{equation}
\ca{E}(A:B)[O]=\lim_{n_e \to 1}  \ca{E}^{(n_e)}(A:B)[O].
\end{equation}


\paragraph{Reflected Entropy}

\begin{figure}
    \centering
    \includegraphics[width = \textwidth]{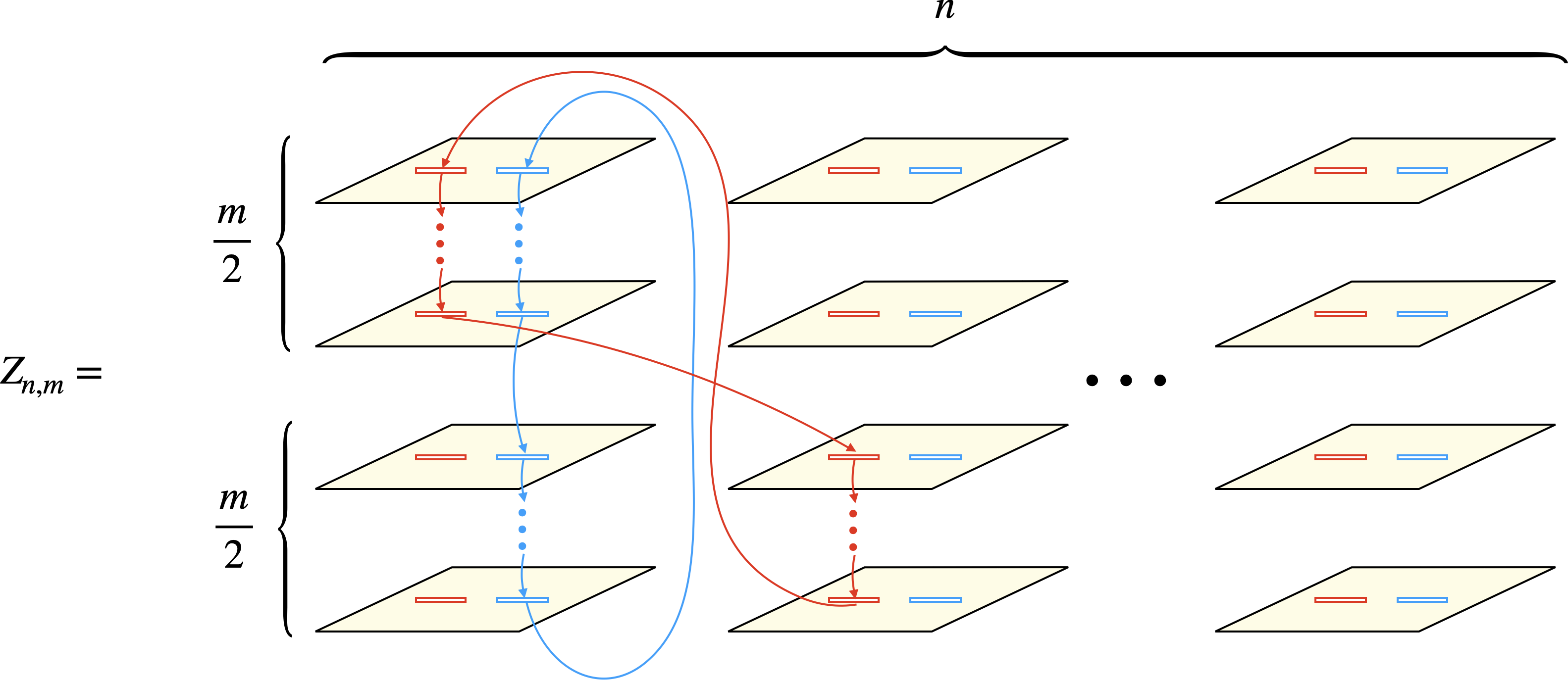}
    \caption{The replica trick for reflected entropy requires two replica numbers, $n$ and $m$. The copies are glued cyclically in $m$ for region $B$ for each value of $n$ (blue line). For region $A$, the gluing mixes different $n$'s in the order shown by the red line. Note that it is necessary for $m$ to be even.}
    \label{SR_replica}
\end{figure}

The path integral representation of the reflected entropy was formulated in Ref.~\cite{2019arXiv190500577D}.
In holographic CFTs, this measure for the excited state (\ref{eq:state}) has already studied in Refs.~\cite{2019arXiv190706646K, 2019arXiv190906790K}.
The R\'enyi reflected entropy is expressed in terms of the ratio of multi-sheeted partition functions (Fig.~\ref{SR_replica}) as
\begin{equation}\label{eq:RenyiRE}
S_R^{(n,m)}(A:B)[O]= \frac{1}{1-n} \log \Tr \left( \left[ \rho_{A A^*}\right]^n\right)=
	\fr{1}{1-n} \log
	\fr{Z_{n,m}}
	{\pa{ Z_{1,m}}^n }.
\end{equation}
where we define $\rho_{AA^*}$ in terms of (\ref{eq:purification}) by
\begin{equation}
\rho_{AA^*} = \tr_{BB^*} \ket{\rho_{AB}^{m/2}}\bra{\rho_{AB}^{m/2}}.
\end{equation}
For disjoint intervals, the replica partition function may be computed by six-point correlation functions
\begin{equation}\label{eq:dRenyi}
Z_{n,m}\equiv \Braket{\sigma_{g_A}(u_1)\sigma_{g_A^{-1}}(v_1)  {O^{\otimes mn}}(w_1,\bar{w}_1)  \dg{{O^{\otimes mn}}}(w_2,\bar{w}_2)   \sigma_{g_B}(u_2) \sigma_{g_B^{-1}}(v_2) }_{\text{CFT}^{\otimes mn}}.
\end{equation}
To avoid unnecessary technicalities, we do not show the precise definition of the twist operators $\sigma_{g_A}$ and $\sigma_{g_B}$ (which can be found in Ref.~\cite{2019arXiv190500577D}) because in this paper, we only use the conformal dimensions and OPEs
\begin{align}
&h_{\sigma_{g_A}}=h_{\sigma_{g_A^{-1}}}=h_{\sigma_{g_B}}=h_{\sigma_{g_B^{-1}}}=\fr{cn}{24} \pa{m-\fr{1}{m}}  (= n h_m)   ,\\
& \sigma_{g_A^{-1}} \times \sigma_{g_B} =  \sigma_{g_n} + \dots, \quad   
\end{align}
where the twist operator $\sigma_{g_n}$ is just the usual twist operator $\sigma_n$ used previously for mutual information and negativity.
The reflected entropy is defined by the von Neumann limit
\begin{equation}
S_R(A:B)[O]=\lim_{n,m \to1} S_R^{(n,m)}(A:B)[O].
\end{equation}
It is crucial that we continue $m$ from the \textit{even} integers to one, similar to the replica trick for negativity.

\subsection{Regge limit}

In general, correlations measured between disjoint intervals, $A$ and $B$, for
the state \eqref{state}
are given in terms of six-point functions, which are too complicated to evaluate in complete generality.
The key to making these computations tractable is to make use of recent progress in understanding the structure of Virasoro conformal blocks \cite{2018JHEP...01..115K,2019JHEP...01..025K,2019JHEP...05..212C,2019arXiv190502191K}, specifically, Regge limit asymptotics.
Roughly, the limit $\epsilon \to 0$ of the correlation measures after the excitation can be evaluated in terms of a single conformal block as \cite{2019arXiv190706646K, 2019arXiv190906790K, 2019arXiv190502191K}
\newsavebox{\boxpa}
\sbox{\boxpa}{\includegraphics[width=180pt]{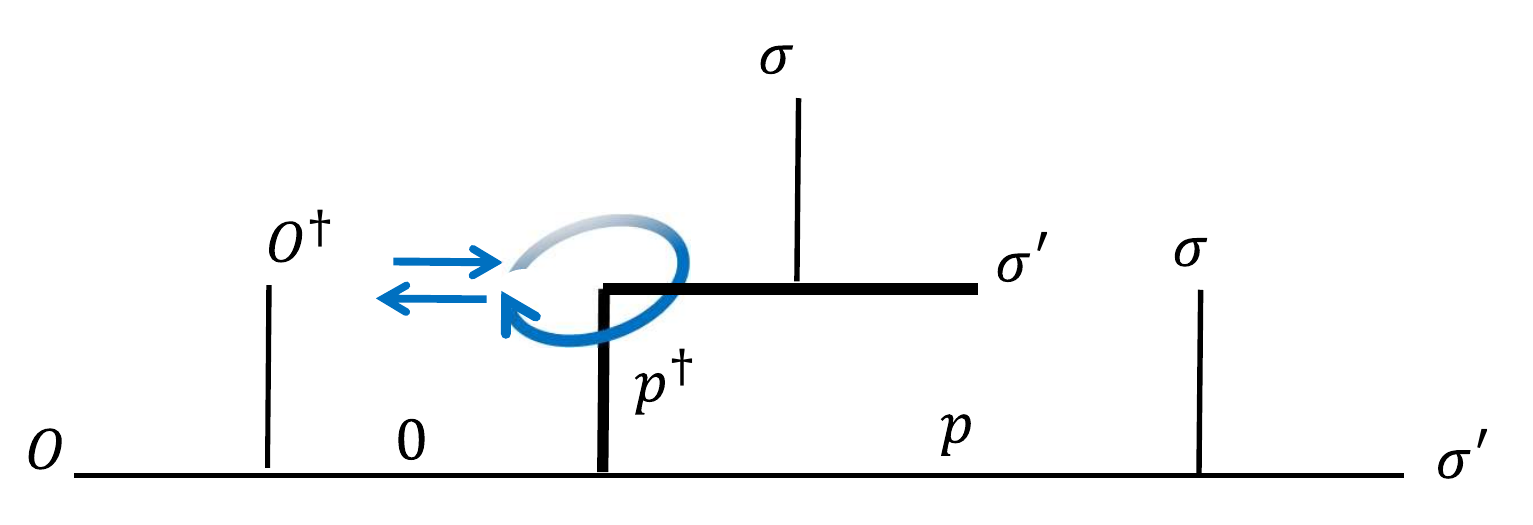}}
\newlength{\paw}
\settowidth{\paw}{\usebox{\boxpa}} 

\newsavebox{\boxpb}
\sbox{\boxpb}{\includegraphics[width=180pt]{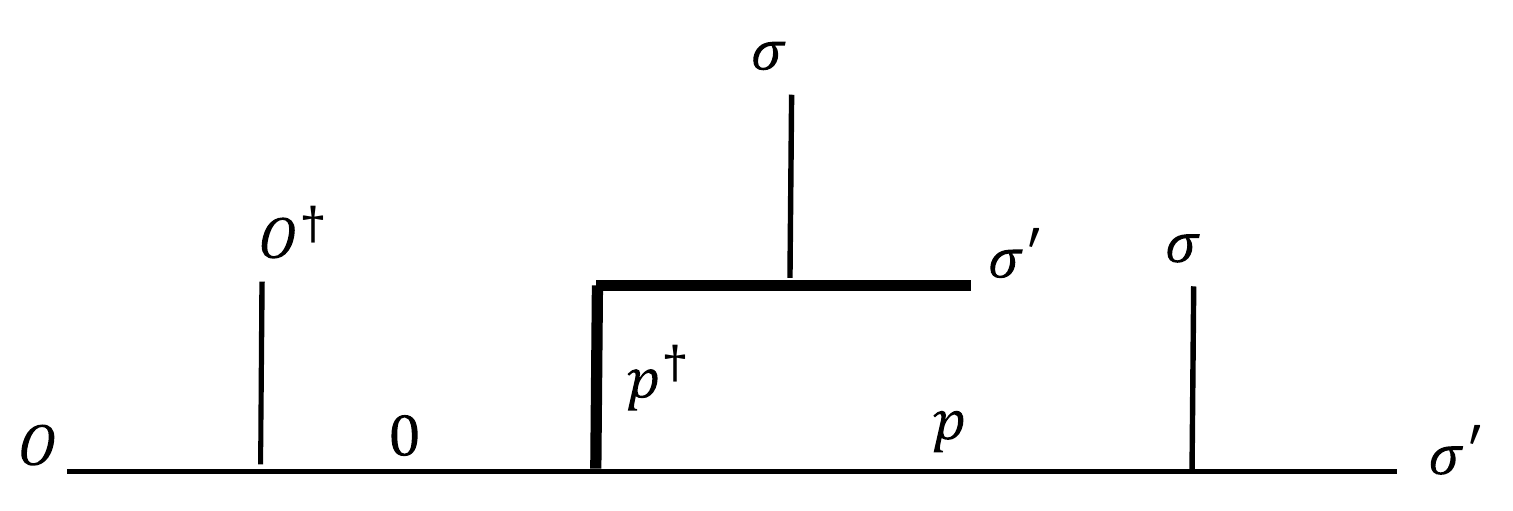}}
\newlength{\pbw}
\settowidth{\pbw}{\usebox{\boxpb}} 

\begin{equation}\label{eq:localchannel}
{(C_{\sigma \sigma' p})}^2 \parbox{\paw}{\usebox{\boxpa}} \times \overline{\parbox{\pbw}{\usebox{\boxpb}} },
\end{equation}
where $C_{\sigma \sigma' p}$ is the OPE coefficient and the arrow implies the monodromy of $\dg{O}$ around the exchanged operator $p$, which is the operator with the lowest dimension in the OPE between $\sigma$ and $\sigma'$.
This monodromy effect is encapsulated by the {\it monodromy matrix}, and is completely determined by the four external operators $ \{ O, \dg{O}, p, \dg{p} \}$ and intermediate state $\{ 0 \}$. 
Therefore, we can study the local quench protocols for disjoint intervals, even when this corresponds to $n(>4)$-point correlation functions.
This is the reason why we can reveal the precise dynamics of the correlation measures.



\section{Integrable conformal field theories}
\label{RCFT}
We begin our analysis with rational conformal field theories. These theories are special in that they have a finite number of primary fields. In some sense, this renders the theories integrable. For simplicity, we consider minimal models but we can straightforwardly generalize our result to general RCFTs\footnote{Generic RCFTs require understanding the monodromy matrix that corresponds to the extended symmetry algebras.}. 

We will see that the crucial difference between RCFTs and generic $c>1$ CFTs is that in RCFTs, the dominant contribution to the Regge asymptotics comes from the vacuum instead of a nontrivial primary.

\subsection*{Mutual Information}

Here we consider the difference of mutual information between the local quench state and the vacuum state, which is denoted by $\D$ in the following. 
As a concrete example, we consider the setup described in Fig.~\ref{fig:setup}. Namely, we set our subregions to $A=[u_1,v_1]$ and $B=[u_2,v_2]$ with $0< u_2<-v_1<-u_1<v_2$. 
We focus on the time region $-v_1<t<-u_1$, where the nontrivial entanglement between $A$ and $B$ is created. 

In order to evaluate the mutual information after a local quench, let us consider the Regge limit of the correlator corresponding to the third term of the definition (\ref{eq:MIdef}),
\begin{equation}
\Braket{\sigma_{n}(u_1)\sigma_{n}^{-1}(v_1)  O^{\otimes n}(w_1,\bar{w}_1)  \dg{{O^{\otimes n}}} (w_2,\bar{w}_2)   \sigma_{n}(u_2) \sigma_{n}^{-1}(v_2) }_{\text{CFT}^{\otimes n}}.
\end{equation}
The lowest dimension operator in the $\sigma_n$ and $\sigma_n^{-1}$ OPE is the vacuum. Therefore, this six-point function can be approximated by

\newsavebox{\boxpc}
\sbox{\boxpc}{\includegraphics[width=180pt]{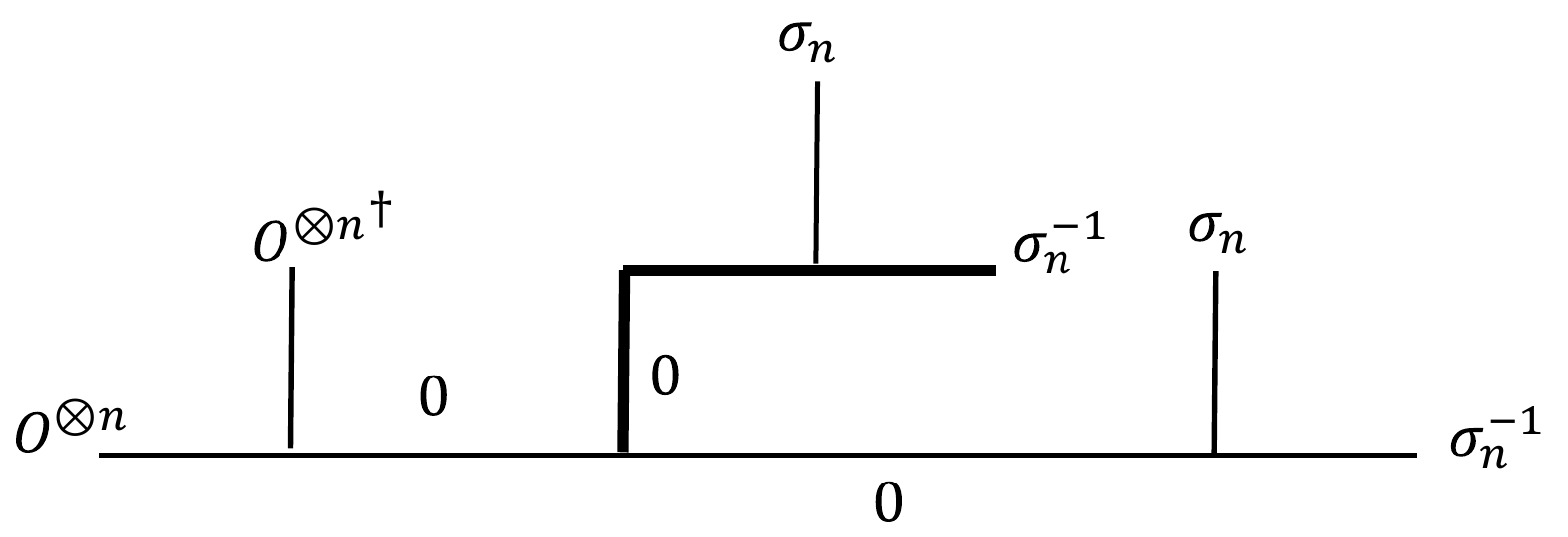}}
\newlength{\pcw}
\settowidth{\pcw}{\usebox{\boxpc}} 

\newsavebox{\boxpd}
\sbox{\boxpd}{\includegraphics[width=180pt]{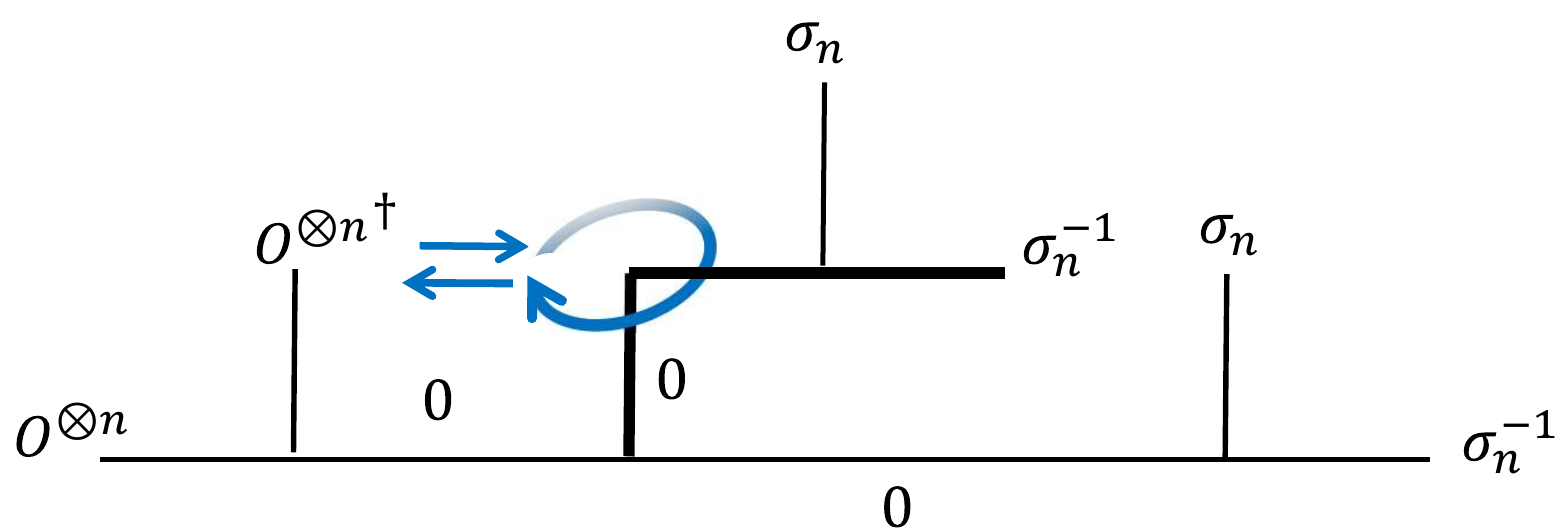}}
\newlength{\pdw}
\settowidth{\pdw}{\usebox{\boxpd}} 

\begin{equation}
\parbox{\pdw}{\usebox{\boxpd}} \times \overline{\parbox{\pcw}{\usebox{\boxpc}}} .
\end{equation}
The monodromy acts on the vacuum which does not change the conformal block.
As a result, we obtain
\begin{equation}
S(AB)[O]=S(AB)[\bb{I}].
\end{equation}
On the other hand, a non-trivial monodromy effect can be found in the four-point correlators that correspond to the first and the second terms in (\ref{eq:MIdef})
\begin{equation}
\Braket{\sigma_{n}(u_1)\sigma_{n}^{-1}(v_1)  O^{\otimes n}(w_1,\bar{w}_1)  \dg{{O^{\otimes n}}} (w_2,\bar{w}_2)  }_{\text{CFT}^{\otimes n}}.
\end{equation}
 The Regge limit of the function corresponding to $S(A)[O]$ (or $S(B)[O]$) can be approximated by
\newsavebox{\boxpf}
\sbox{\boxpf}{\includegraphics[width=120pt]{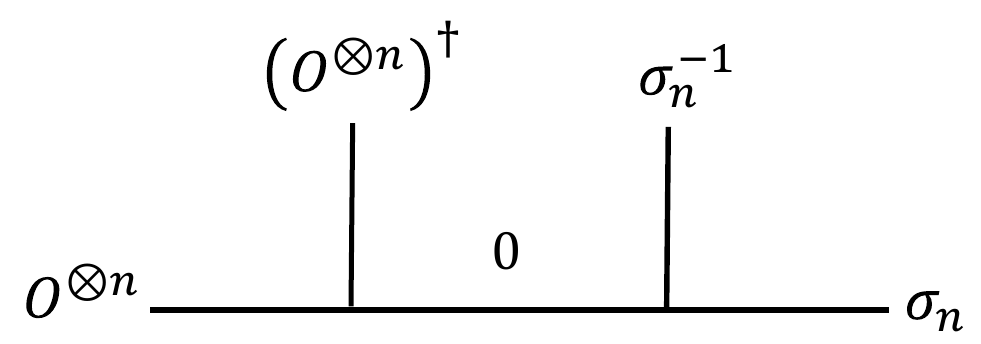}}
\newlength{\pfw}
\settowidth{\pfw}{\usebox{\boxpf}} 

\newsavebox{\boxpg}
\sbox{\boxpg}{\includegraphics[width=120pt]{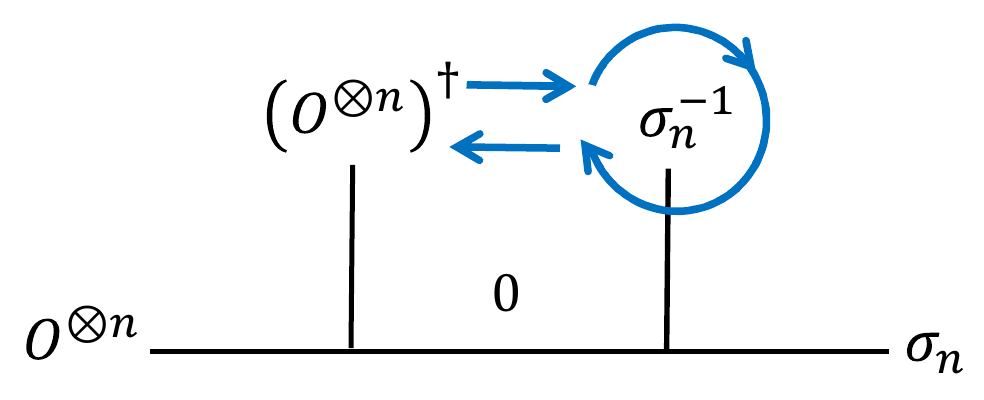}}
\newlength{\pgw}
\settowidth{\pgw}{\usebox{\boxpg}} 

\begin{equation}
 \parbox{\pgw}{\usebox{\boxpg}} \times  \overline{\parbox{\pfw}{\usebox{\boxpf}}}.
\end{equation}
The effect of the monodromy is given as
\begin{equation}
 {{\textbf{ M}}^{(n)}}_{0, 0}[O]    \parbox{\pfw}{\usebox{\boxpf}}.
\end{equation}
That is, the dominant contribution comes from the vacuum. As a result, the remaining term in the difference between the excited state and the vacuum is only the constant $ {{\textbf{ M}}^{(n)}}_{0, 0}[O] $.
This $(0,0)$-element of the monodromy matrix is related to the quantum dimension of $O$, just like the fusion matrix. Thus, we can express the growth of the mutual information in terms of the quantum dimension as
\begin{equation}
\begin{aligned}
\D I(A:B)[O]&=\left\{
    \begin{array}{ll}
  0 
		,   & \text{if } t<-v_1 ,\\ \\
  2\log d_O 
		,   & \text{if } -v_1<t<-u_1,\\ \\
  0
		,   & \text{if } -u_1<t .
    \end{array}
  \right.\\
\end{aligned}
\end{equation}
In the pure state limit ($B \to \overline{A}$), the mutual information should match twice the entanglement entropy $S(A)[O]$. Our result is consistent with this fact.

We are able to generalize this result to any R\'enyi entropy by using the explicit form of the monodromy matrix, 
\begin{equation}\label{eq:MIinRCFT}
\begin{aligned}
\D I^{(n)}(A:B)[O]&=\left\{
    \begin{array}{ll}
  0 
		,   & \text{if } t<-v_1 ,\\ \\
  2\log d_O 
		,   & \text{if } -v_1<t<-u_1,\\ \\
  0
		,   & \text{if } -u_1<t .
    \end{array}
  \right.\\
\end{aligned}
\end{equation}

\subsection*{Logarithmic Negativity}
To obtain the logarithmic negativity, we need the Regge limit of the following correlator,
\begin{equation}
\Braket{\sigma_{n}(u_1)\sigma_{n}^{-1}(v_1)  O^{\otimes n}(w_1,\bar{w}_1)  \dg{{O^{\otimes n}}} (w_2,\bar{w}_2)   \sigma_{n}^{-1}(u_2) \sigma_{n}(v_2) }_{\text{CFT}^{\otimes n}}.
\label{neg_corr}
\end{equation}
We have to analytically continue $n$ from even integers to one.
This analytic continuation requires us to take special care of the effect of the orbifolding.
We introduce the following notations:
\begin{equation}
O^{\otimes n}=O^{\otimes n/2}_{(1)} \otimes O^{\otimes n/2}_{(2)},
\end{equation}
where the subscript 1(2) implies that the operator acts like the local primary operator $O$ on the odd (even) numbered
sheets and the identity on the even (odd) sheets. Analogously, this is how the double twist field (for even $n$) decomposes
\begin{equation}
\sigma_n^2= \sigma_{n/2}^{(1)}  \otimes\sigma_{n/2}^{(2)} .
\end{equation}
The crucial point is that $\{ O^{\otimes n/2}_{(1)}, \sigma_{n/2}^{(1)} \}$ do not interact with $\{ O^{\otimes n/2}_{(2)}, \sigma_{n/2}^{(2)} \}$. Therefore, the component of the (three-point) conformal block decouples into two parts, for example,
\begin{equation}\label{eq:dec}
\braket{   \sigma_{n/2}^{(1)}  \otimes\sigma_{n/2}^{(2)} |O^{\otimes n/2}_{(1)} \otimes O^{\otimes n/2}_{(2)}  |  \sigma_{n/2}^{(1)}  \otimes\sigma_{n/2}^{(2)}  } = 
\braket{  \sigma_{n/2}^{(1)}   |O^{\otimes n/2}_{(1)} |   \sigma_{n/2}^{(1)}  } 
\braket{  \sigma_{n/2}^{(2)}   |O^{\otimes n/2}_{(2)}  |   \sigma_{n/2}^{(2)}  } .
\end{equation}
In other words, the vacuum sector in the conformal block decomposition can be perfectly decomposed into two manifolds, one composed of even sheets and the other composed of odd sheets.

Using the Regge limit, we can approximate \eqref{neg_corr} by a single conformal block
\newsavebox{\boxpi}
\sbox{\boxpi}{\includegraphics[width=180pt]{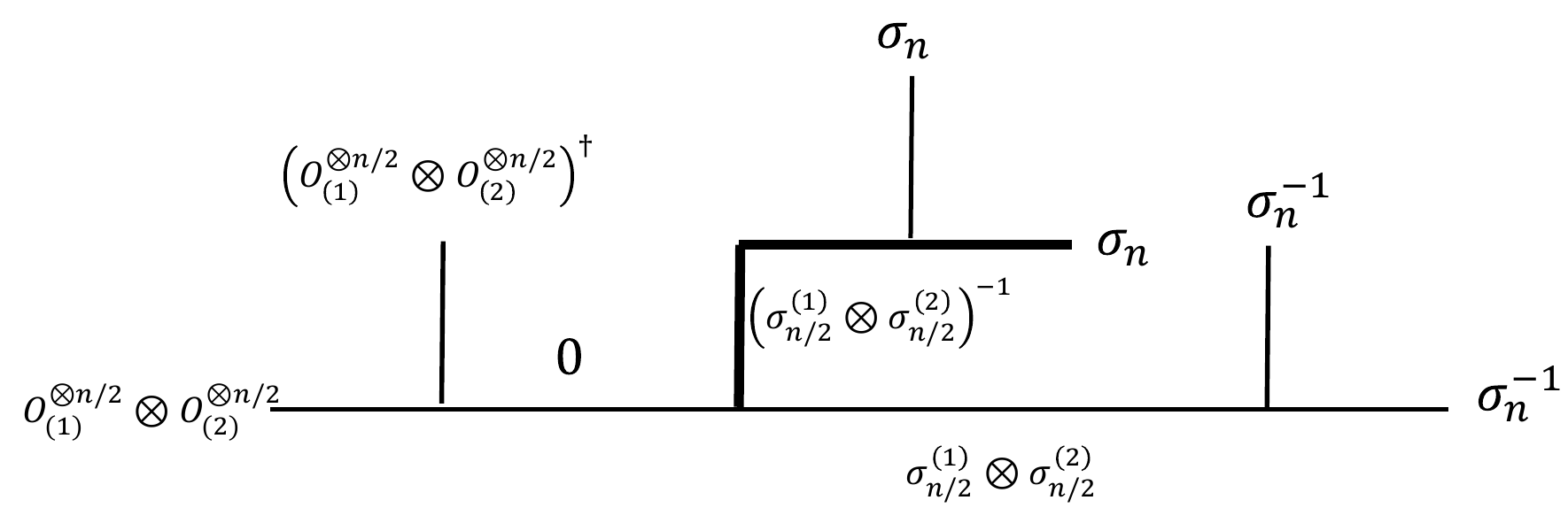}}
\newlength{\piw}
\settowidth{\piw}{\usebox{\boxpi}} 
\newsavebox{\boxpj}
\sbox{\boxpj}{\includegraphics[width=180pt]{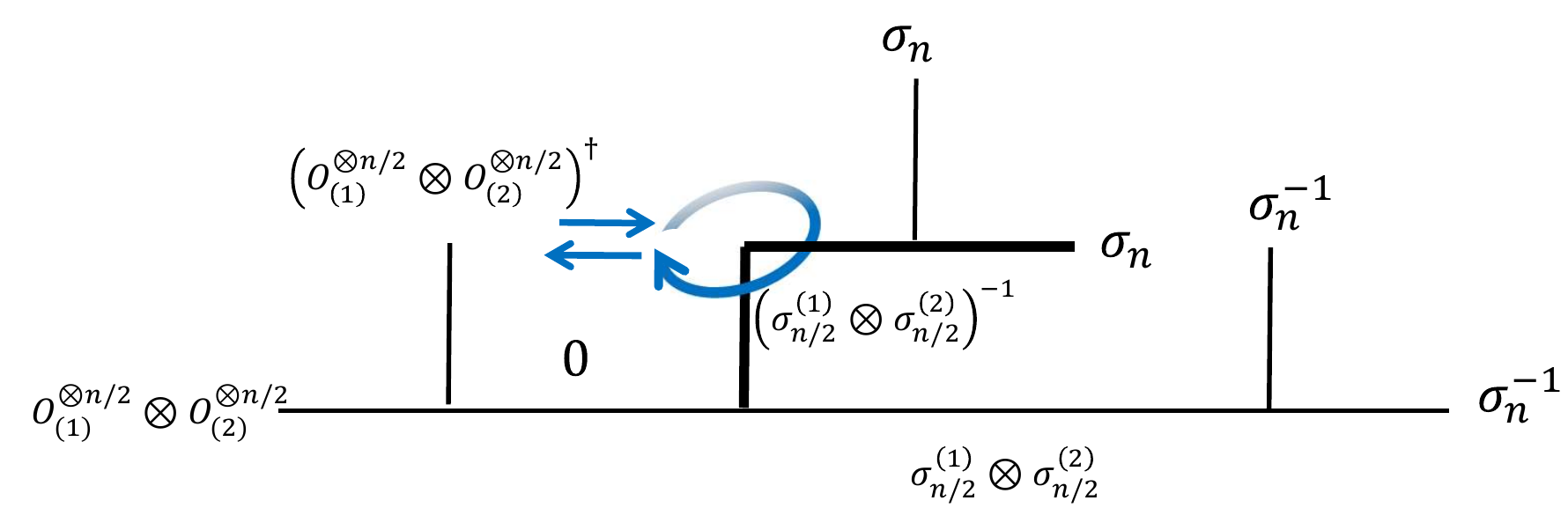}}
\newlength{\pjw}
\settowidth{\pjw}{\usebox{\boxpj}} 
\begin{equation}
(C_n)^4 \parbox{\pjw}{\usebox{\boxpj}} \times  \parbox{\piw}{\usebox{\boxpi}},
\end{equation}
where $C_n$ is the OPE coefficient $\Braket{\sigma_n |\pa{\sigma_{n/2}^{(1)}\otimes\sigma_{n/2}^{(2)}}^{-1} | \sigma_n }$.
As mentioned in the review, the Regge singularity between $O^{\otimes n}$ and $\dg{\pa{O^{\otimes n}}}$ is only determined by four external operators
\begin{equation}
\biggl\{   \pa{\sigma_{n/2}^{(1)}  \otimes\sigma_{n/2}^{(2)}}    , \pa{\sigma_{n/2}^{(1)}  \otimes\sigma_{n/2}^{(2)}}^{-1},   O^{\otimes n}    , \dg{\pa{O^{\otimes n}}}  
\biggr\}
\end{equation}
and its intermediate state (which is now given as the vacuum). Moreover, this part is decomposed into two parts due to (\ref{eq:dec}).
For the same reason as for the mutual information, the corresponding Regge limit is 
\newsavebox{\boxpl}
\sbox{\boxpl}{\includegraphics[width=130pt]{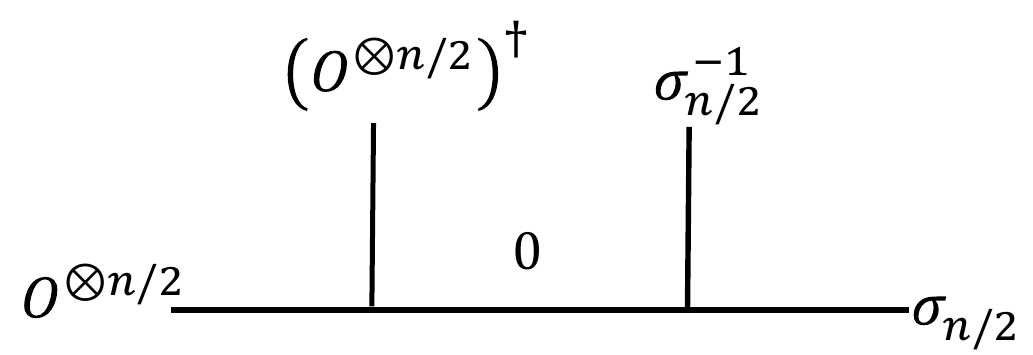}}
\newlength{\plw}
\settowidth{\plw}{\usebox{\boxpl}} 
\begin{equation}
\pa{ {{ \textbf{M}}^{(n/2)}}_{0, 0}[O]     \parbox{\plw}{\usebox{\boxpl}} }^2,
\end{equation}
which gives the following behavior for the negativity
\begin{equation}\label{eq:NinRCFT}
\begin{aligned}
\D \ca{E}(A:B)[O]&=\left\{
    \begin{array}{ll}
  0 
		,   & \text{if } t<-v_1 ,\\ \\
  \log d_O
		,   & \text{if } -v_1<t<-u_1,\\ \\
  0
		,   & \text{if } -u_1<t .
    \end{array}
  \right.\\
\end{aligned}
\end{equation}
It is known that in the pure state limit ($B \to \overline{A}$), the negativity reduces to the R\'enyi entropy at index $1/2$,
\begin{equation}
\ca{E}(A,\overline{A})=S^{(1/2)}(A).
\end{equation}
As shown in Ref.~\cite{2014arXiv1403.0702H}, the $n$-th R\'enyi entropy after a local quench does not depend on $n$,
\begin{equation}
\D S^{(n)}(A)[O]=\log d_O, \ \ \  \text{if } -v_1<t<-u_1.
\end{equation}
One can see that our calculation utilizing the Regge limit of conformal blocks is consistent with this fact.

\subsection*{Reflected Entropy}

For the reflected entropy, we need to compute the following correlation function
\begin{equation}
Z_{n,m}\equiv \Braket{\sigma_{g_A}(u_1)\sigma_{g_A^{-1}}(v_1)  {O^{\otimes mn}}(w_1,\bar{w}_1)  \dg{{O^{\otimes mn}}}(w_2,\bar{w}_2)   \sigma_{g_B}(u_2) \sigma_{g_B^{-1}}(v_2) }_{\text{CFT}^{\otimes mn}}.
\label{sr_corr}
\end{equation}
To explain various subtleties, we introduce the following notation,
\begin{align}
    O^{\otimes m_e n} = O^{\otimes n}_{(0)} \otimes \dots \otimes O^{\otimes n}_{(m_e/2)} \otimes \dots,
\end{align}
where the subscript labels the replica copy in the $m_e$ direction where the operator acts.
In the limits $m_e \to 1$ and $\e \to 0$, \eqref{sr_corr} can be approximated by
\newsavebox{\boxRE}
\sbox{\boxRE}{\includegraphics[width=180pt]{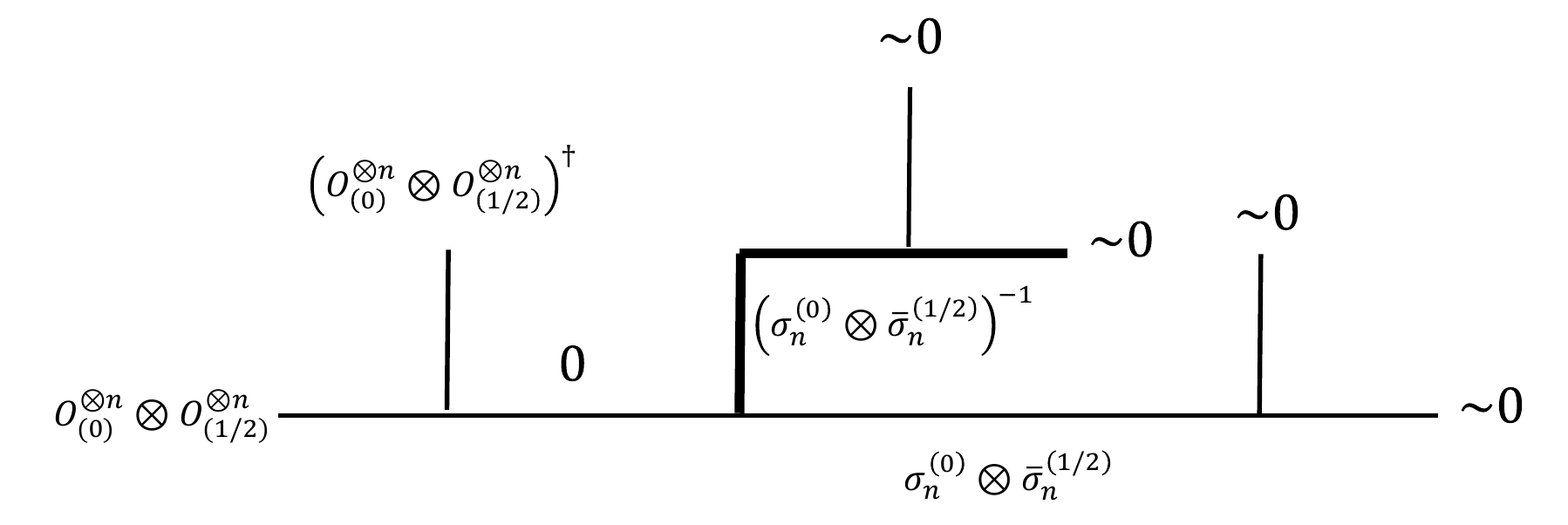}}
\newlength{\REw}
\settowidth{\REw}{\usebox{\boxRE}} 
\newsavebox{\boxREm}
\sbox{\boxREm}{\includegraphics[width=180pt]{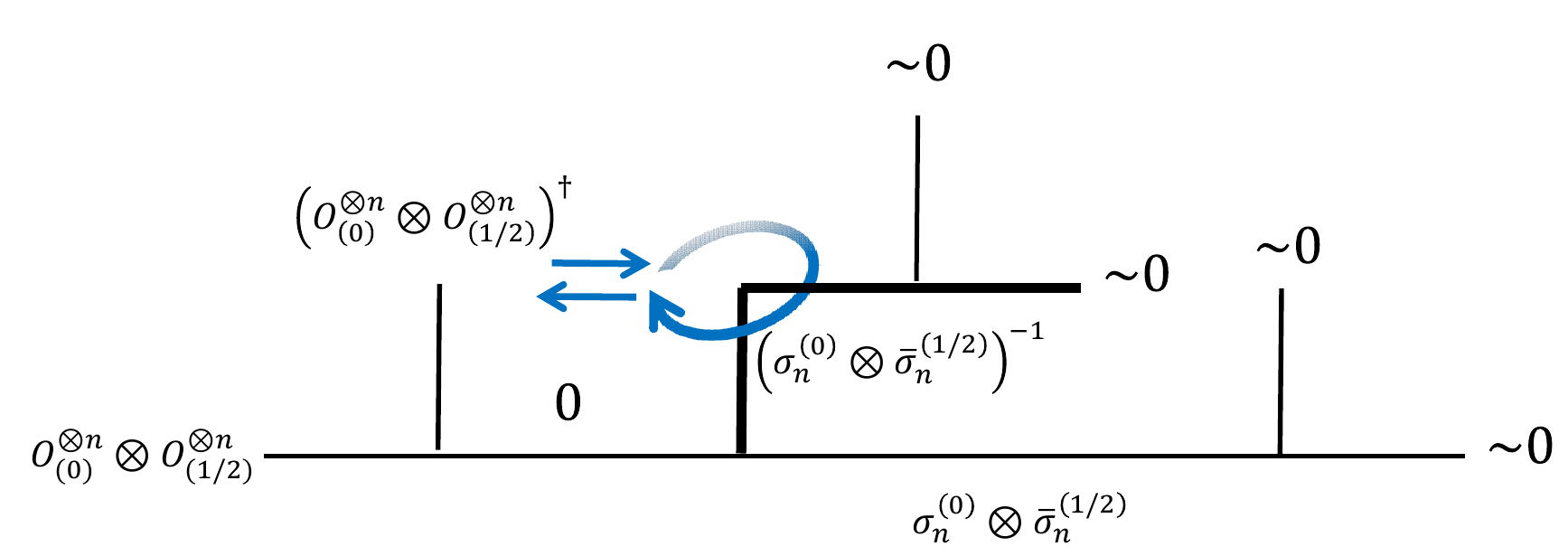}}
\newlength{\REmw}
\settowidth{\REmw}{\usebox{\boxREm}} 
\begin{equation}
(\ca{C}_n)^4 \parbox{\REmw}{\usebox{\boxREm}} \times  \parbox{\REw}{\usebox{\boxRE}},
\end{equation}
where $\ca{C}_n$ is the OPE coefficient $\lim_{m_e \to 1} \Braket{\sigma_{m_e} |\pa{\sigma_{n}^{(0)}\otimes \bar{\sigma}_{n}^{(1/2)}}^{-1} | \sigma_{m_e} }$ and $\sim 0$ means a state with very small conformal dimension. Similarly to (\ref{eq:dec}), the three-point block in the above 6-point conformal block can be decomposed into two parts,
\begin{equation}
\braket{  \sigma_n^{(0)}  \otimes  \bar{\sigma}_n^{(1/2)}  |  O^{\otimes n}_{(0)} \otimes   O^{\otimes n}_{(1/2)}   | \sigma_n^{(0)}  \otimes \bar{\sigma}_n^{(1/2)}  } = 
\braket{  \sigma_n^{(0)}   |  O^{\otimes n}_{(0)} |   \sigma_n^{(0)}  } 
\braket{  \sigma_n^{(1/2)}   |  O^{\otimes n}_{(1/2)} |   \sigma_n^{(1/2)}  }.
\end{equation}
As a result, we obtain a {\it square} of the monodromy matrix $\pa{{{\textbf{M}}^{(n)}}_{0, 0}[O]}^2$ in the Regge limit correlator. 

The canonical purification for the reflected entropy also leads to a subtle squaring in the normalization factor.
The analytic continuation $m_e \to 1$ reduces to \cite{PhysRevLett.123.131603, 2019arXiv190906790K}
\begin{align}
    \lim_{m_e \rightarrow 1} O^{\otimes m_en} = O^{\otimes n}_{(0)} \otimes O^{\otimes n}_{(1/2)}.
\end{align}
A similar squaring can be found in the twist operators for reflected entropy
\begin{align}
   \lim_{m_e \rightarrow 1}  \sigma_{g_A^{-1} g_B} = \sigma_n^{(0)} \otimes \bar{\sigma}_n^{(1/2)}.
\end{align}
As a result from this squaring, we obtain the squared correlation function,
\begin{align}
\lim_{m_e \rightarrow 1} Z_{1,m_e}
=  \Braket{O(w_1,\bar{w}_1)\dg{O}(w_2,\bar{w}_2)}^{2}.
\end{align}

In summary, we have
\begin{equation}
\label{RCFT_SR}
\begin{aligned}
\D S_R^{(n)}(A:B)[O]&=\left\{
    \begin{array}{ll}
  0 
		,   & \text{if } t<-v_1 ,\\ \\
  2\log d_O 
		,   & \text{if } -v_1<t<-u_1,\\ \\
  0
		,   & \text{if } -u_1<t .
    \end{array}
  \right.\\
\end{aligned}
\end{equation}
Note the if we naively evaluate the reflected entropy by regarding the block as the Virasoro block, we obtain an incorrect result, ${{\textbf{M}}^{(n)}}_{0, 0}[O]$, insted of $\pa{{{\textbf{M}}^{(n)}}_{0, 0}[O]}^2$. Therefore, we have to take care of the squaring of this conformal block. We refer the interested reader to Ref.~\cite{2019arXiv190906790K} for more technical details.

\subsection*{Correlation web}
Having completed all calculations for RCFTs, we would like to check our conjectures relating each measure. From \eqref{eq:MIinRCFT} and \eqref{eq:NinRCFT}, we can confirm \eqref{cal_al_prop}. Note, however, that from \eqref{eq:MIinRCFT} and \eqref{eq:NinRCFT}, we can rule out the more general validity of the relation from Ref.~\cite{2020arXiv200105501K} which showed the negativity to be proportional to all R\'enyi mutual informations for a specific class of quantum quenches that did not involve any local operator insertions. Together with the results from this paper and Refs.~\cite{2018arXiv180909119A, 2020arXiv200105501K}, we believe that the statement of \eqref{cal_al_prop} for generic integrable theories is on very strong footing. 

We also are able to confirm consistency of both \eqref{sr_I_conj} and \eqref{ln_sr_conj} in integrable systems from \eqref{eq:NinRCFT} and \eqref{RCFT_SR}. Next, we show how all of these relations except for \eqref{ln_sr_conj} break down when we eliminate the integrability of the CFT.

\section{Chaotic conformal field theories}
\label{pureCFT}

We progress to $c>1$ CFTs with finite twist gap. While these theories are believed to be generic, we do not know of any explicit constructions. This class of theories displays chaotic properties even for finite central charge \cite{Asplund2015a, 2019arXiv190502191K, 2020JHEP...01..175K,2020arXiv200105501K}. It will be convenient for us to introduce Liouville coordinates,
\begin{equation}
c=1+6Q^2, \ \ \ \ \ Q=b+\fr{1}{b}, \ \ \ \ \ h_i=\a_i(Q-\a_i).
\end{equation}

\subsection*{Mutual Information}
For the same reason as in RCFTs, the contribution from $S(AB)[O]$ is equal to $S(AB)[\mathbb{I}]$ during the time regime of interest and can therefore be neglected. The nontrivial contributions to the change in mutual information will come from $S(A)[O]$ and $S(B)[O]$.
The Regge limit of the relevant correlation functions can be approximated by
\begin{equation}
 \parbox{\pgw}{\usebox{\boxpg}} \times  \overline{\parbox{\pfw}{\usebox{\boxpf}}}.
\end{equation}

Although this is similar to the evaluation in RCFTs, the monodromy effect is dramatically different due to the absence of vacuum exchange in the cross-channel
\newsavebox{\boxph}
\sbox{\boxph}{\includegraphics[width=120pt]{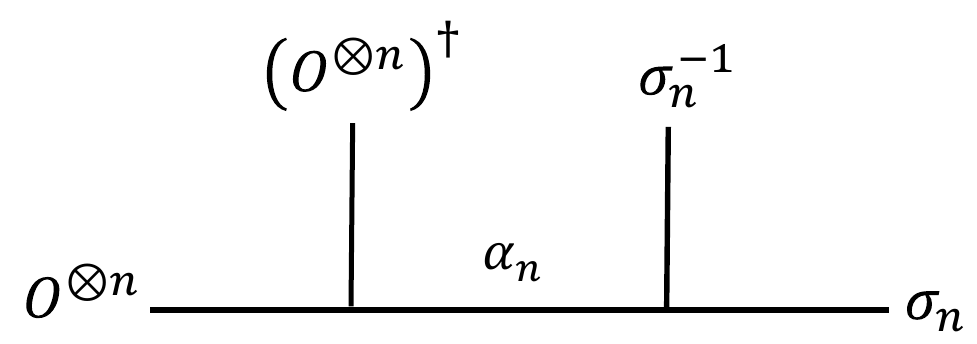}}
\newlength{\phw}
\settowidth{\phw}{\usebox{\boxph}} 
\begin{equation}
 {{\textbf{ M}}^{(n)}}_{0, \a_n}[O]    \parbox{\phw}{\usebox{\boxph}} \ar{\e \to 0}  (2i\e)^{h_{\a_n}} {{ \textbf{M}}^{(n)}}_{0, \a_n}[O] \rho_{\sigma_n \sigma_n^{-1} \a_n}(z_i)  ,
\end{equation}
where the constant $ {{\textbf{ M}}^{(n)}}_{0, \a_n}[O] $ is related to the monodromy matrix\footnote{
More precisely, this constant is given by the coefficient of the first order pole of the monodromy matrix \cite{2019arXiv190502191K}.
In RCFTs, the monodromy transformation is given by the summation over a discrete spectrum. On the other hand, the monodromy transformation in pure CFTs is given by an integral over a continuous spectrum. Nevertheless, the leading order in the Regge limit is approximated by a single residue, $\text{Res}(-2 \pi i  {{\textbf{ M}}^{(n)}}_{0, \a}; \a = \a_n)$. We express this constant as ${{\textbf{ M}}^{(n)}}_{0, \a_n}$ to avoid this cumbersome expression.}
and $\a_n$ is the minimal Liouville momentum in the Regge OPE between $\sigma_n$ and $\sigma_n^{-1}$. 
In the von Neumann limit, this minimal value reduces to $h_{\a_n}=2h_n$.
We abbreviate the three point block as $\rho_{\sigma_n \sigma_n^{-1} \a_n}(z_i)$, which is given by
\begin{equation}
\rho_{\sigma_n \sigma_n^{-1} \a_n}(t,u_1,v_1)=\pa{\fr{(v_1-u_1)}{(t-u_1) (t-v_1)}}^{h_{\a_n}} (v_1-u_1)^{-2h_n}  .
\end{equation}
Consequently, we obtain the R\'enyi entanglement entropy as
\begin{equation}
\begin{aligned}
\D S^{(n)}(A)[O]=\fr{h_{\a_n}}{n-1} \log\fr{ (t-u_1) (t-v_1)  }{2i \e (v_1-u_1)}+\fr{1}{1-n} \log  \br{   {{ \textbf{M}}^{(n)}}_{0,\a_n}[O]  }.
\end{aligned}
\end{equation}
Hence, the $n$-th R\'enyi mutual information is given by
\begin{equation}\label{eq:MIresult}
\begin{aligned}
\D I^{(n)}(A:B)[O]
&=\fr{h_{\a_n}}{n-1} \log\fr{ (t-u_1) (t-v_1)  }{2i \e (v_1-u_1)}
+\fr{h_{\a_n}}{n-1} \log\fr{ (t-u_2) (t-v_2)  }{2i \e (v_2-u_2)}
\\
&\qquad
+\fr{2}{1-n} \log  \br{ {{\textbf{M}}^{(n)}}_{0,\a_n}[O]  }\\
&\equiv
\fr{h_{\a_n}}{n-1} \log\fr{ g^{\text{MI}}(z_i)  }{\pa{2i\e}^2}
+\fr{2}{1-n} \log  \br{ {{\textbf{M}}^{(n)}}_{0,\a_n}[O]  },
\end{aligned}
\end{equation}
where we define
\begin{equation}
g^{\text{MI}}(z_i) \equiv \fr{ (t-u_1) (t-v_1)  }{ (v_1-u_1)}\fr{ (t-u_2) (t-v_2)  }{ (v_2-u_2)}.
\end{equation}
The von Neumann limit is
\begin{equation}
\begin{aligned}
\D I(A:B)[O]=\fr{c}{6} \ \log\fr{ g^{\text{MI}}(z_i)  }{\pa{2i\e}^2} +  \lim_{n \to 1} \fr{2}{1-n} \log  \left[{   {{\textbf{M}}^{(n)}}_{0,\a_n}[O] }\right].
\end{aligned}
\end{equation}
Some comments are in order. (i) The mutual information grows logarithmically without bound, sharply contrasting the RCFT result. Unbounded logarithmic growth cannot be described by the quasi-particle picture and and is a signature of multipartite entanglement generation. (ii) The second term corresponding the leading residue of the $Vir^n/\mathbb{Z}_n$ monodromy matrix is constant in time. In Ref.~\cite{2019arXiv190502191K}, it was argued that this term may simplify in the limit of large central charge, in which case $\textbf{M}^{(n)}_{0,\a_n}[O]\simeq \left(\textbf{M}_{0,\a_n}[O]\right)^n$, where $\textbf{M}_{0,\a_n}[O]$ corresponds to the leading residue of the Virasoro monodromy matrix. In the von Neumann limit, this term can then be evaluated, matching gravitational calculations
\begin{align}
    \lim_{n \to 1, c\rightarrow \infty} \fr{2}{1-n} \log  \left[ -2i {{\textbf{M}}^{(n)}}_{0,\a_n}[O] \right] = 4\pi \sqrt{\frac{c}{6}\left(h_{O}-\frac{c}{24} \right)},
\end{align}
which is equal to twice the Cardy entropy.
\subsection*{Logarithmic Negativity}

In a similar way as the above computation for mutual information, the growth of the negativity is encapsulated in the square of the four-point conformal block as
\newsavebox{\boxpk}
\sbox{\boxpk}{\includegraphics[width=130pt]{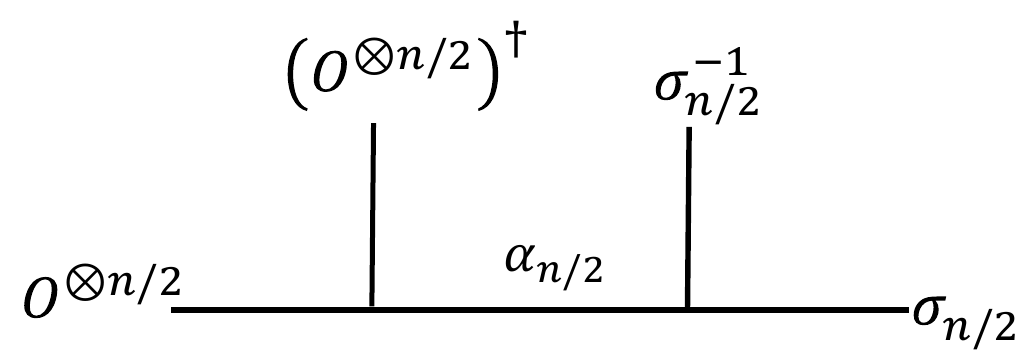}}
\newlength{\pkw}
\settowidth{\pkw}{\usebox{\boxpk}} 
\begin{equation}\label{eq:partN}
\pa{ {{\textbf{M}}^{(n/2)}}_{0, \a_n}[O]    \parbox{\pkw}{\usebox{\boxpk}} }^2.
\end{equation}
More precisely, the $\epsilon \to 0$ (Regge) limit leads to
\newsavebox{\boxpm}
\sbox{\boxpm}{\includegraphics[width=150pt]{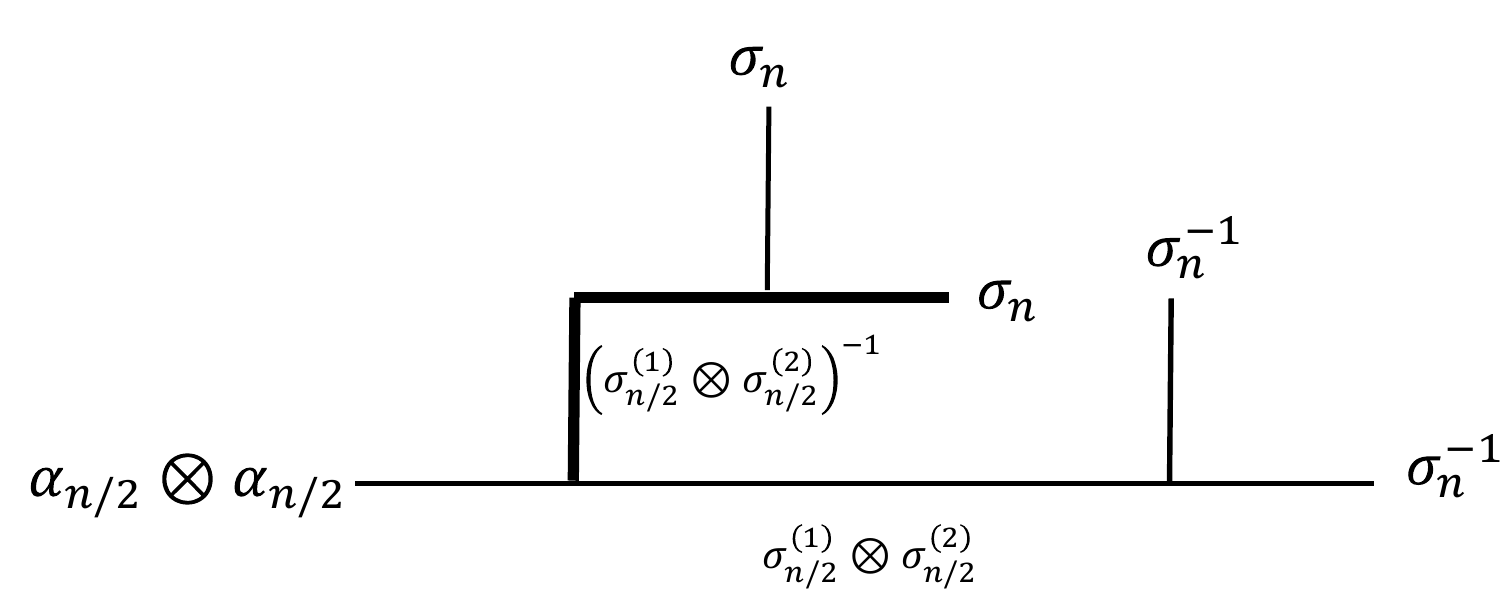}}
\newlength{\pmw}
\settowidth{\pmw}{\usebox{\boxpm}} 
\begin{equation}\label{eq:Ncomp}
\begin{aligned}
&\pa{ 2i \e  }^{2h_{\a_{n/2}}-2nh_O} \pa{ {{\textbf{M}}^{(n/2)}}_{0, \a_{n/2}}[O] }^2  \parbox{\pmw}{\usebox{\boxpm}} \\
&\equiv   \pa{ 2i \e  }^{2h_{\a_{n/2}}-2nh_O} \pa{ {{\textbf{M}}^{(n/2)}}_{0, \a_{n/2}}[O] }^2 g^{\ca{E}^{(n/2)}}(z_i)^{-h_{\a_{n/2}}},
\end{aligned}
\end{equation}
where we abbreviate the five-point conformal block as $ g^{\ca{E}^{(n/2)}}(z_i)^{-h_{\a_{n/2}}}$, which has the following asymptotics,
\begin{equation}
g^{\ca{E}^{(n/2)}}(z_i) \ar{ \abs{u_2-v_1} ,\abs{u_1-v_2}\ll t   } \pa{-t}^2.
\end{equation}
As a result, we obtain
\begin{equation}\label{eq:Nresult}
\begin{aligned}
\D \ca{E}(A:B)[O]= -h_{\a_{1/2}}  \log\fr{g^{\ca{E}^{(1/2)}}(z_i)}{ \pa{ 2i \e}^2  }+  2\log  \br{   {{\textbf{M}}^{(1/2)}}_{0,\a_{1/2}}[O] }.
\end{aligned}
\end{equation}
According to Ref.~\cite{2019arXiv190502191K}, we can expect that the explicit form of $h_{\a_n}$ can be expressed by

\begin{equation}
h_{\a_n}=2\b_n(Q-2\b_n),
\end{equation}
where $\b_n(Q-\b_n)\equiv h_n$. In particular, the classical value at $n=\fr{1}{2}$ is
\begin{equation}
h_{\a_{1/2}} \ar{c \to \infty} \left(-\fr{5+\sqrt{10}}{12} \right)c.
\end{equation}

In this section, we have made the crucial, non-rigorous assumption that we may use the dominant \textit{Virasoro} conformal block. In reality, we have an extended $\mathbb{Z}_n$ symmetry due to the replica trick. It would be more precise to take the dominant $Vir^n/ \mathbb{Z}_n$ block, though, unfortunately, these are not well understood. While we expect the precise answer to be either the same or very close to \eqref{eq:Nresult}, we are able to argue in Appendix \ref{app_vir_vs_zn} that \eqref{eq:Nresult} should be at least a lower bound on the negativity. Even if we treat this as a bound, it is quite interesting, particularly gravitationally, as we explore in Section \ref{backreaction_sec}.

\subsection*{Reflected Entropy}

For the reflected entropy, we obtain a similar expression in the Regge limit to that for the negativity as
\begin{equation}
 \pa{ 2i \e  }^{2h_{\a_n}- 4nh_O } \pa{ {{\textbf{M}}^{(n)}}_{0, \a_n}[O] }^2 g^{\ca{E}^{(n)}}(z_i)^{-h_{\a_n}},
\end{equation}
where $ g^{\ca{E}^{(n)}}$ is defined below \eqref{eq:Ncomp}.
As a result, we obtain
\begin{equation}\label{eq:REresult}
\begin{aligned}
\D S(A:B)_R[O]= \fr{c}{6} \log\fr{g^{\ca{E}^{(1)}}(z_i)}{ \pa{ 2i \e}^2  }+  \lim_{n \to 1} \fr{2}{1-n}\log  \br{   {{\textbf{M}}^{(n)}}_{0,\a_{n}}[O] },
\end{aligned}
\end{equation}
where the von Neumann limit of $g^{\ca{E}^{(n)}}(z_i)$ is calculated by the global block as \cite{2019arXiv190906790K}
\begin{equation}
g^{\ca{E}^{(n)}}(z_i) \ar{n \to 1}  \fr{(t+u_1)(t+u_2)(t+v_1)(t+v_2)}{(u_2-v_1)(u_1-v_2)} .
\end{equation}
Note that the expression for the R\'enyi reflected entropy is given by 
\begin{equation}\label{eq:REresult2}
\begin{aligned}
\D S^{(n)}_R(A:B)[O]= \fr{h_{\a_n}}{n-1}  \log\fr{g^{\ca{E}^{(n)}}(z_i)}{ \pa{ 2i \e}^2  }+  \fr{2}{1-n}\log  \br{   {{\textbf{M}}^{(n)}}_{0,\a_{n}}[O] }.
\end{aligned}
\end{equation}

\subsection*{Correlation web}

We now return to the collection of relations between the various correlation measures. 
From \eqref{eq:Nresult} and \eqref{eq:REresult2}, one can confirm the relation between the negativity and the R\'enyi reflected entropy \eqref{ln_sr_conj} for pure CFTs.
From (\ref{eq:MIresult}) and (\ref{eq:Nresult}), one can find
\begin{equation}
\begin{aligned}
\fr{1}{2}\D I^{(1/2)}(A:B)[O]-\D \ca{E}(A:B)[O]=-h_{\a_{(1/2)}} \log \fr{g^{\text{MI}}(z_i)}{g^{\ca{E}^{(1/2)}}(z_i)}.
\end{aligned}
\end{equation}
While we expect this difference to be non-zero in general, we have several comments suggesting their similarity. (i)
The dependencies on the regulator $\e$ for $\D \ca{E}$ and $\D I^{(1/2)}$ match one another. (ii)
According to Ref.~\cite{2019arXiv190502191K}, the constant ${\textbf{M}}$ is related to the mass of the black hole in the bulk side.
This implies that the topological contribution to  $\D \ca{E}$ is the same as that of $\D I^{(1/2)}$.
(iii)
In a special limit $ \abs{u_2-v_1}, \abs{u_1-v_2} \ll t $, we have
\begin{equation}
g^{\text{MI}}(z_i) = g^{\ca{E}^{(1/2)}}(z_i),
\end{equation}
which leads to
\begin{equation}\label{eq:con}
\D \ca{E}=\D \fr{1}{2}I^{(1/2)}.
\end{equation}
This is consistent with the expectation for RCFTs but is a nontrivial statement for pure CFTs.

Finally, we can see from \eqref{eq:MIresult} and \eqref{eq:REresult2} that \eqref{sr_I_conj} breaks down for pure CFTs for the same reason that \eqref{cal_al_prop} breaks down. We stress that we are directly able to identify the mechanism of the breakdown of the quasi-particle picture. Namely, for pure CFTs, the OPE in the Regge limit does not contain the vacuum state. This fact alone destroys the quasi-particle picture. In contrast, the vacuum state is able to propogate for RCFTs and the quasi-particle picture is restored. 

\section{Holographic negativity and backreaction}
\label{backreaction_sec}

In Ref.~\cite{2019PhRvD..99j6014K}, it was proposed that the holographic dual of logarithmic negativity is the area of a tensionful entanglement wedge cross-section ($\equiv E_W$). Due to the finite tension of this surface, there is nontrivial backreaction in the bulk AdS spacetime. While the full proposal was formally proven in AdS$_3$/CFT$_2$ \cite{PhysRevLett.123.131603}, this backreaction is very difficult to compute in practice as it involves solving Einstein's equations with codimension-two sources. Fortunately, it was observed that the backreaction may be accounted for in sufficiently symmetric states and subsystem configurations by an overall proportionality constant \cite{2019PhRvD..99j6014K, PhysRevLett.123.131603}
\begin{align}
    \mathcal{E} = \frac{3}{2}E_W.
    \label{LN_EW_approx}
\end{align}
This is equivalent to taking the dominant \textit{global} conformal block while ignoring the descendent states \cite{2020JHEP...01..031K}. \eqref{LN_EW_approx} has been confirmed to precisely compute the logarithmic negativity for e.g.~single intervals at zero and finite temperature, adjacent and disjoint intervals at zero and finite temperature, and the thermofield double state.

The local operator quench is certainly not a symmetric state due to the explicit breaking of translation symmetry, so a priori, it would seem that one would need to go through the full gravitational computation with cosmic branes. However, it is interesting and computationally relevant to ask how well (\ref{LN_EW_approx}) still approximates the holographic negativity, a much milder gravitational computation.

For this purpose, we study the holographic dual of a local operator quench, which may be understood as a falling particle in AdS \cite{2013JHEP...05..080N}. In Poincar\'e coordinates
\begin{align}
    ds^2 = \frac{-dt^2+ dx^2 +dz^2}{z^2},
\end{align}
the falling particle has trajectory
\begin{align}
    z^2 - t^2 = \epsilon^2,
\end{align}
where $\epsilon$ corresponds to the smearing of the operator in the CFT language and is related to the energy. The massive particle backreacts on the vacuum AdS geometry as it falls radially. This backreaction can be accurately modeled by boosting a black hole geometry. See Ref.~\cite{2013JHEP...05..080N} for a more detailed discussion.

We will compare the area of the entanglement wedge cross section in these coordinates to the large-$c$ limit of our results for pure CFTs (\ref{eq:Nresult}).
The gravitational computation is identical, up to proportionality, to that of Refs.~\cite{2019arXiv190706646K, 2019arXiv190906790K}. For convenience, we reproduce the results for $-v_1 < t< -u_1$
\begin{align}
    E_W &= \frac{c}{12}\log\left[ \frac{4(t+ u_1)(t+u_2)(t+v_1)(t+ v_2)}{\epsilon^2(u_2-v_1)(u_1- v_2) }\left(\frac{\sinh \pi \bar{\gamma}}{\bar{\gamma}} \right)^2 \frac{1 + \sqrt{\frac{(v_1 - u_1)(v_2-u_2)}{(u_2-u_1)(v_2-v_1)}}}{1 - \sqrt{\frac{(v_1 - u_1)(v_2-u_2)}{(u_2-u_1)(v_2-v_1)}}}\right],
\end{align}
where $\bar \gamma \equiv \sqrt{ \frac{24 \bar{h}}{ c} - 1}$.
To isolate the time dependence and compare to the CFT result, we are interested in the limit $v_1, u_2 \ll t\ll u_1,v_2$ where
\begin{align}
   \Delta E_W = \frac{c}{6} \log\left[ \frac{t}{\epsilon }\left(\frac{\sinh \pi \bar{\gamma}}{\bar{\gamma}} \right) \right].
   \label{delta_EW_limit}
\end{align}
We are able to find that the time dependent piece of the negativity, while logarithmic in $t$, has a different overall coefficient than (\ref{LN_EW_approx}). Instead of a proportionality factor of $3/2$, we find that it is corrected to approximately
\begin{align}
    \frac{\mathcal{E}}{E_W} \simeq {5-\sqrt{10}}  \simeq 1.84 > 3/2.
    \label{extra_backreaction_eq}
\end{align}
The backreaction from the particle has a net positive effect on the area of the tensionful entanglement wedge cross section.
Note that the constant part 
in Eq.\ \eqref{delta_EW_limit}
corresponds to the Bekenstein-Hawking entropy (see Ref.~\cite{2019arXiv190502191K}).
If we assume equation (5.55) of Ref.~\cite{2019arXiv190502191K}, then we obtain the same entropy in the above negativity from the monodromy matrix.

\section{Lattice models}

\label{lattice}

So far, we have only been concerned with conformally invariant theories. In this section, we will argue that these special systems are able to detect certain universal aspects of integrable and chaotic dynamics. We argue this by numerically simulating local operator quenches in lattice models. For free fermions, we are able to implement finite scaling analysis to confirm the consistency of (\ref{cal_al_prop}), but are only able to provide preliminary evidence for interacting integrable systems and chaotic spin chains. 

\subsection*{Free fermions}

\begin{figure}
    \centering
    \includegraphics[width = .48\textwidth]{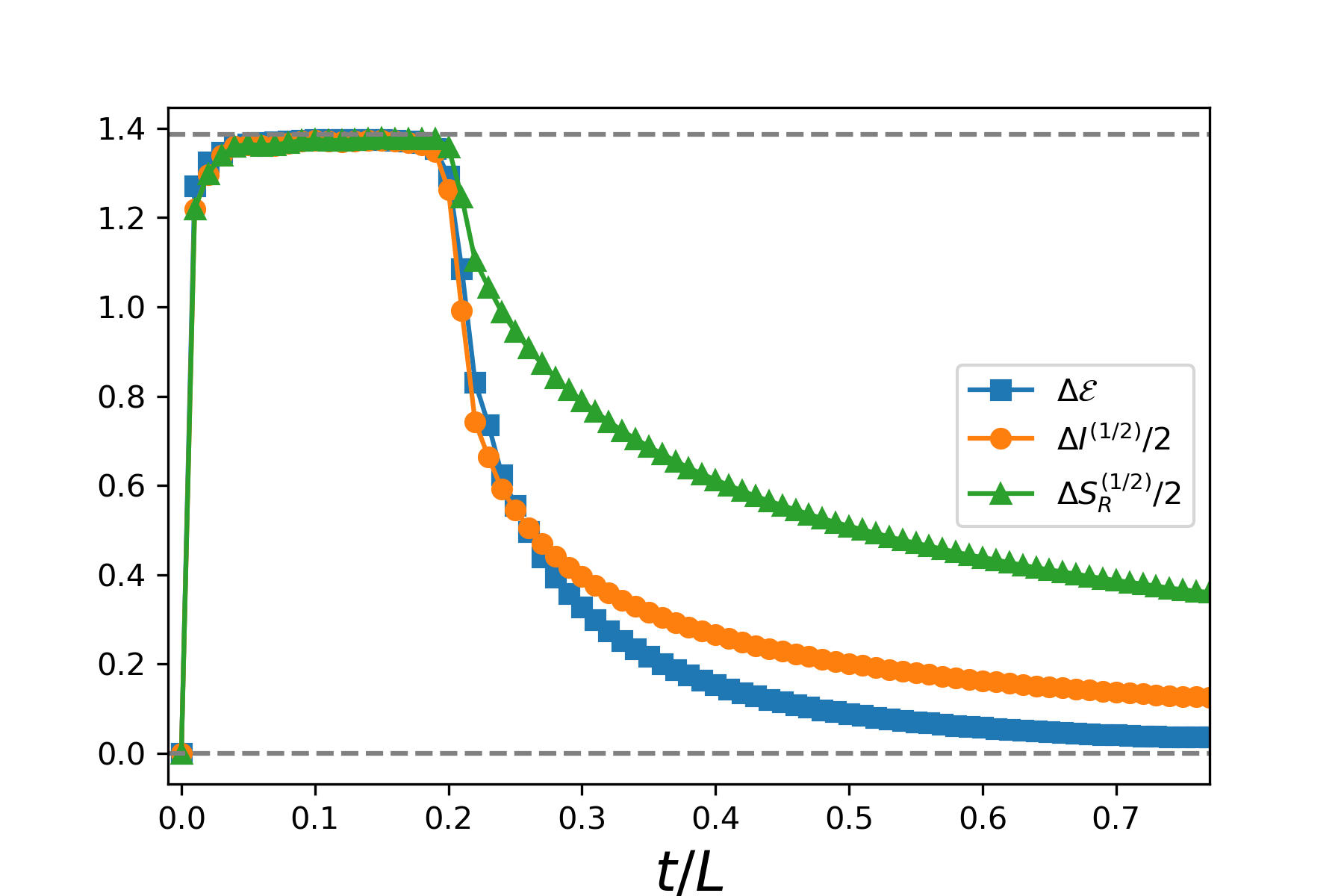}
    \includegraphics[width = .48\textwidth]{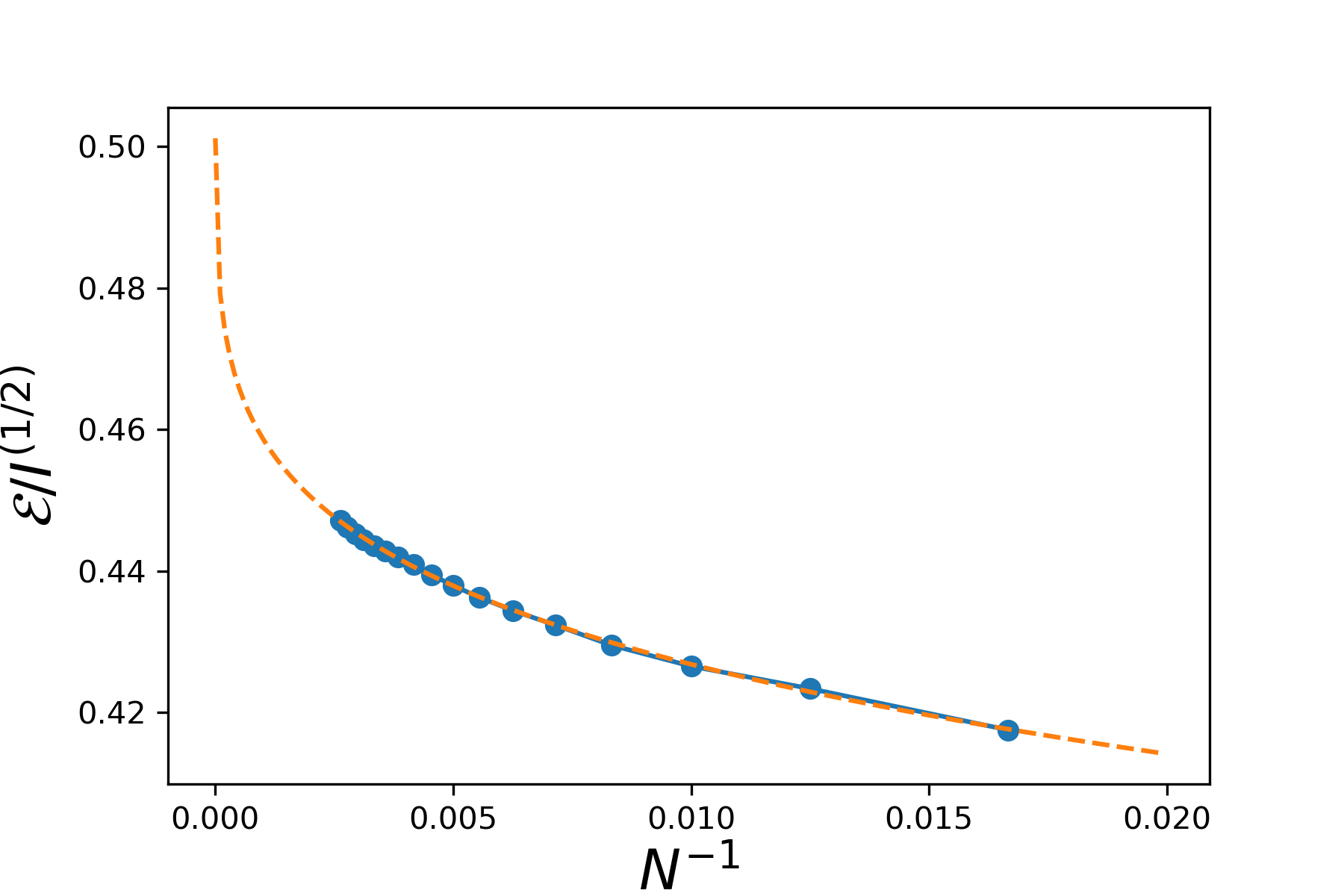}
    \caption{Comparison between negativity, mutual information, and reflected entropy in free fermion systems after a local excitation by the fermion parity operator $(-1)^F$. Left: $\Delta \mathcal{E}$ (blue) $\Delta I^{(1/2)}/2$ (orange) $\Delta S_R^{(1/2)}/2$ (green), $N = 750$, $l_1 = l_2 = N/5$. The dotted line is at $2\log 2$. Right: $l_1 = l_2 = N/5$, $t_f = 3N/10$. fit: $a N^{-b}+c$ with $a = -0.216  (-0.2202, -0.2118)$, $b =      0.1901  (0.1342, 0.246)$, $c =      0.5168  (0.4925, 0.541)$. The value of $c$ is consistent with (\ref{cal_al_prop}). Similar finite-size effect deviations between $\mathcal{E}$ and $I^{(1/2)}/2$ were found after a global quench in Ref.~\cite{2018arXiv180909119A}.}
    \label{ff_fig}
\end{figure}

For free fermions, we use the tight-binding Hamiltonian 
\begin{equation}
    \hat{H}= \frac{1}{2}\left(\sum^N_{i}  \hat{c}_i^\dagger \hat{c}_{i+1} + h.c.
    \right)
\end{equation}
and apply anti-periodic boundary conditions. For simplicity, we use adjacent intervals of lengths $l_1$ and $l_2$ and insert the operator at their interface. The results are shown in Fig.~\ref{ff_fig} where we see that $\mathcal{E}$, $I^{(1/2)}/2$, and  $S_R^{(1/2)}/2$\footnote{
The entanglement negativity 
can be efficiently computed 
for fermionic Gaussian states 
by using the correlator method
\cite{2017PhRvB..95p5101S}.
Similarly,
we were able to compute the reflected entropy and its R\'enyi counterparts efficiently by using the correlator method where the correlation matrix for the purified state was recently constructed in Ref.~\cite{2020arXiv200309546B}.} are nearly identical at early times, but diverge once the local quasi-particle excitation has left the subsystems. 
We perform finite scaling analysis in Fig.~\ref{ff_fig} to confirm that this late-time behavior of the quantities converges in the scaling limit. The long-time tail of the correlations is due to ``high-energy quasi-particles,'' a feature also observed in joining quenches \cite{2015PhRvB..92g5109W}.

\subsection*{Spin chains}

While free systems are computationally tractable, they are trivially integrable and in the above case, conformal at low energies. 
We are ultimately interested in interacting systems. 
For this purpose, we compute the negativity and R\'enyi mutual information in the transverse field Ising model
\begin{align}
    H = \sum_i(- Z_iZ_{i+1} + g X_i + hZ_i).
    \label{sc_H}
\end{align}
We probe integrable dynamics by taking $g = 1.0, h = 0$ and chaotic\footnote{Here, by chaotic, we mean that the level statistics of the Hamiltonian mimic those of random matrix theory.} dynamics by taking $g = -1.05, h = 0.5$.
Because this system is interacting, we are limited to small system sizes. However, we are able to make some progress by representing the local quench state as a matrix product state (MPS) \cite{fannes1992,2004cond.mat..7066V,2007JSMTE..08...24H} and evolve in real time using time evolving block decimation (TEBD) \cite{2003PhRvL..91n7902V}. Furthermore, we find the ground state using density matrix renormalization group methods (DMRG) \cite{PhysRevLett.69.2863,2011AnPhy.326...96S}. We use the python package \textit{quimb} \cite{gray2018quimb} to implement the tensor network techniques. The reason we are able to use MPS methods even in this non-equilibrium scenario is because local quenches do not create that much entanglement. While we necessarily increase the bond dimension of the tensors at later times, we are able to simulate large enough systems to find the central features of integrable and chaotic systems. 

For the integrable spin chain, we observe a rapid growth of entanglement that essentially saturates to a constant. This is reminiscent of the RCFT picture where the quasi-particle created by the local quench produces a step function in the correlation measures. Being nonconformal, the dispersion relation of the theory is nontrivial, so the softening of the step function is expected. This may furthermore be attributed to finite-size effects. For the chaotic spin chain, it is interesting that we can observe a logarithmic growth of the entanglement, similar to the pure CFTs. This clearly demonstrates the breakdown of the quasi-particle picture. It is impressive that conformal field theory computations can capture certain core features that distinguish integrable and chaotic spin chains. It would certainly be interesting to further pursue this direction in lattice models to understand how universal these structures truly are.

\begin{figure}
    \centering
    \includegraphics[width = .48\textwidth]{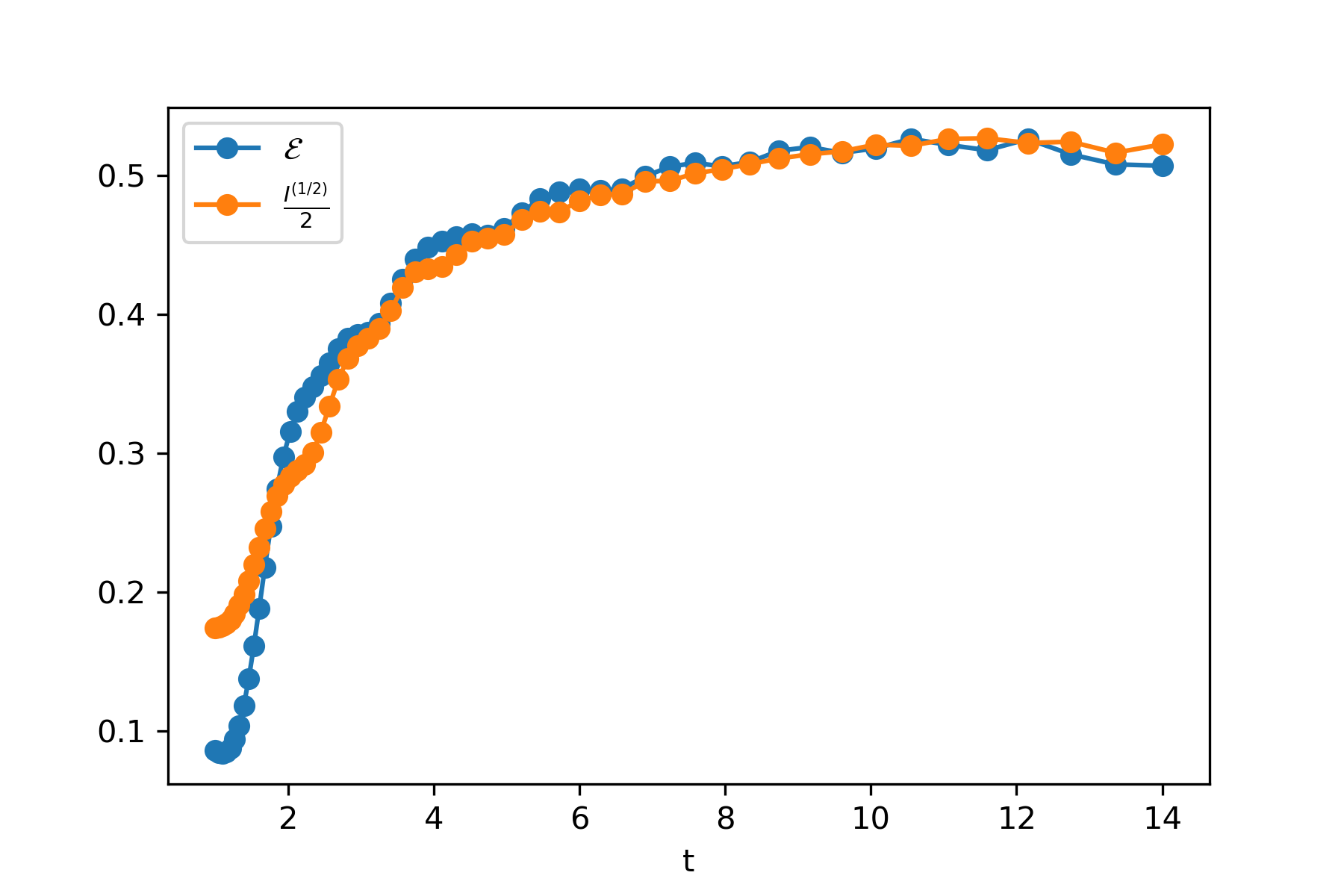}
    \includegraphics[width = .48\textwidth]{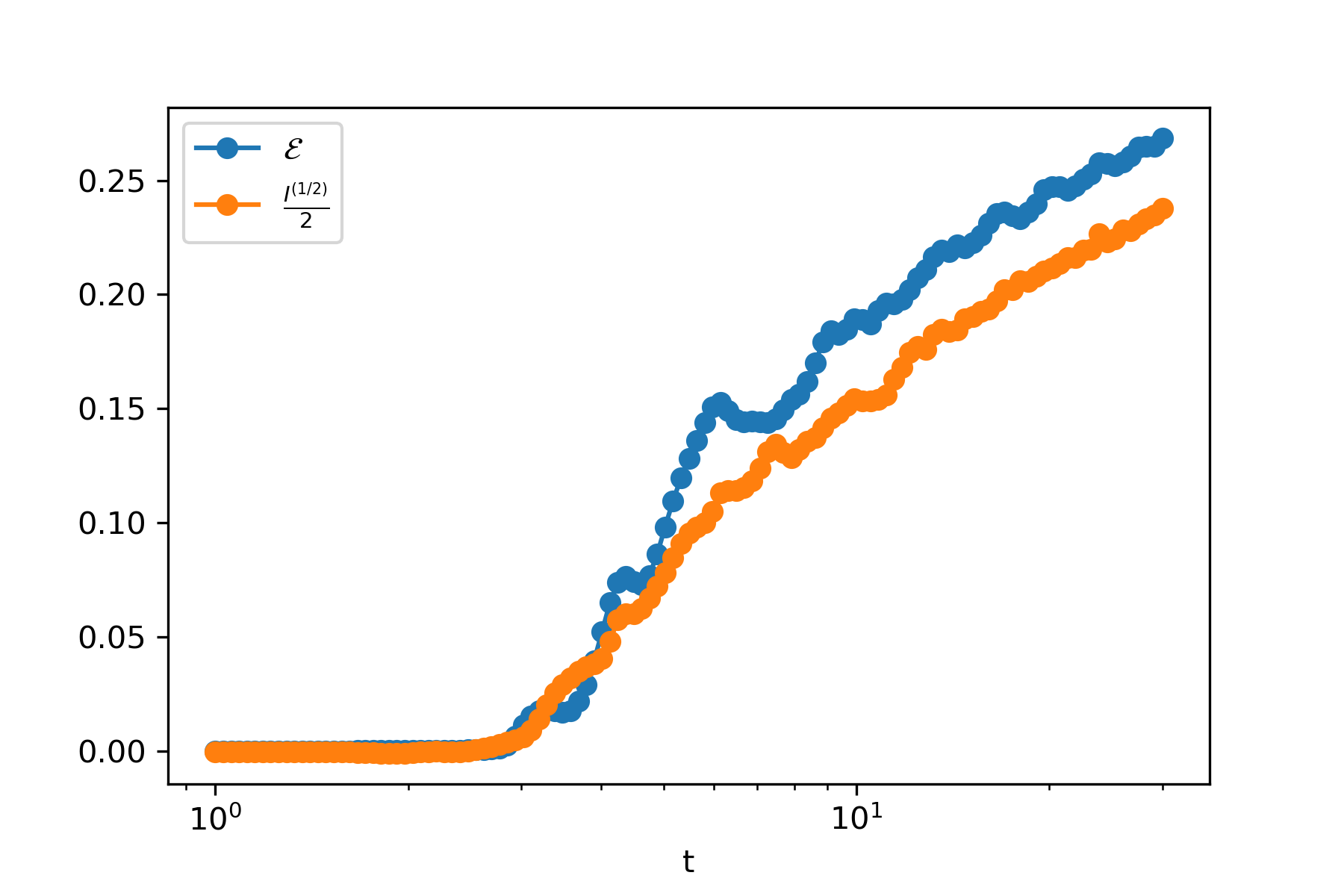}
    \caption{Left: Integrable spin chain. We see a rapid growth to a constant, similar to the behavior for RCFTs, though less sharp due to both finite size effects and a nonlinear dispersion relation. Right: Chaotic spin chain. We observe approximately logarithmic growth of the negativity and R\'enyi mutual information similar to that of pure CFTs. We simulate 31 lattice sites with 5 sites separating ``semi-infinite'' subsystems. The local operator used is a Pauli $Z$.}
    \label{fig:my_label}
\end{figure}

\acknowledgments

We thank Hassan Shapourian and Tadashi Takayanagi for fruitful discussions and comments. We thank the Yukawa Institute for Theoretical Physics at Kyoto University where this work was initiated during the workshop YITP-T-19-03 ``Quantum Information and String Theory 2019.'' YK is supported by a JSPS fellowship. SR is supported by a Simons Investigator Grant from the Simons Foundation.

\appendix

\section{Can we use \texorpdfstring{$Vir$}{TEXT} conformal blocks in place of \texorpdfstring{$Vir^n/\mathbb{Z}_n$}{TEXT} blocks?}
\label{app_vir_vs_zn}

In Section \ref{pureCFT}, we made a key assumption that when evaluating correlation functions in the Regge limit, we can take the dominant Virsaoro conformal block instead of the orbifold block, $Vir^n/\mathbb{Z}_n$. The $Vir$ block includes all contributions from regular descendent states
\begin{align}
    L_{-m_1}^{k_1}\dots L_{-m_j}^{k_j} \ket{h}
\end{align}
where $h$ is some highest weight state, and the $L$'s are the Virasoro generators for the stress tensor of the seed theory. In contrast, the $Vir^n/\mathbb{Z}_n$ block includes all $\mathbb{Z}_n$ symmetric contributions from 
\begin{align}
    L_{-m_1}^{k_1}(l_1)\dots L_{-m_j}^{k_j}(l_j) \ket{h},
\end{align}
where $L(l_p)$ are the Virasoro generators in the $p^{th}$ copy of the theory. It is not clear that naively using the dominant $Vir$ conformal block will reproduce the results for the $Vir^n/\mathbb{Z}_n$ block. In this appendix, we will argue that this assumption is acceptable for the case of the local operator quench. 
More conservatively, we argue the $Vir$ block Regge singularity lower bounds the $Vir^n/\mathbb{Z}_n$ singularity. First, we use heuristic arguments; afterwards, we provide explicit computations that we find moderately convincing.

By taking the $Vir$ block, we are only summing over intermediate states that are ``diagonal'' in the $Vir^n/\mathbb{Z}_n$ block. By diagonal, we mean that in each intermediate state, the exact same Virasoro generators for each of the $n$ copies are applied. These intermediate states are trivially $\mathbb{Z}_n$ symmetric. The ones neglected are $\mathbb{Z}_n$ symmetric but less trivial e.g.~$\sum_{i}^n L_{-k}(i) L_{-k}(i+1) \ket{h}$. These are more complex (boundary) graviton exchanges and can only strengthen the interaction. Thus, the $Vir^n/\mathbb{Z}_n$ block should have at least as strong of a singularity as the $Vir$ block, so we should be able to use this as an interesting lower bound.

Let us now sharpen what we mean by ``strengthen the interaction'' by appealing to the gravitational description of negativity. When we approximate the holographic negativity as $\mathcal{E} = 3/2 E_W$, we are using the dominant \textit{global} conformal block (see Appendix A of Ref.~\cite{2020JHEP...01..031K}). When taking the Virasoro conformal blcok, we have included the ``diagonal'' descendants and consequently increased $\mathcal{E}$ \eqref{extra_backreaction_eq}. Because the other $\mathbb{Z}_n$ symmetric descendants are simply more graviton exchanges, we can expect that these extra gravitons will push the negativity further in the same direction (increase). 

We progress to a more quantitative argument by considering higher-$n$ R\'enyi entropies after a local quench. These may be computed from two independent methods, uniformization maps and twist fields. 
\color{black}

\paragraph*{Uniformization map approach}

We review the computation completed in Ref.~\cite{2019arXiv190502191K}. The integer R\'enyi entropies are computed as
\begin{align}
    \Delta S_n = \frac{1}{1-n} \log \frac{\langle \mathcal{O}(w_1^{(1)}, \bar{w}_1^{(1)}) \mathcal{O}(w_2^{(1)}, \bar{w}_2^{(1)})\dots \mathcal{O}(w_1^{(n)} ,\bar{w}_1^{(n)}) \mathcal{O}(w_2^{(n)}, \bar{w}_2^{(n)})\rangle_{\Sigma_n}}{\langle\mathcal{O}(w_1 ,\bar{w}) \mathcal{O}(w_2 , \bar{w}_2) \rangle_{\Sigma_1}}
\end{align}
where $\Sigma_n$ is the $n$-branched cover of the original manifold and $w^{i}$ are the coordinates on the $i^{th}$ copy.

We then consider the uniformization map that takes $\Sigma_n$ to a single copy of the complex plane
\begin{align}
    z = \left(\frac{w}{i\epsilon + t -l} \right)^{\frac{1}{n}}.
\end{align}
As in the main text, we take $\epsilon \rightarrow 0$, in which case
\begin{align}
&
    \frac{\langle \mathcal{O}(w_1^{(1)}, \bar{w}_1^{(1)}) \mathcal{O}(w_2^{(1)}, \bar{w}_2^{(1)})\dots \mathcal{O}(w_1^{(n)} ,\bar{w}_1^{(n)}) \mathcal{O}(w_2^{(n)}, \bar{w}_2^{(n)})\rangle_{\Sigma_n}}{\langle\mathcal{O}(w_1 ,\bar{w}) \mathcal{O}(w_2 , \bar{w}_2) \rangle_{\Sigma_1}}  
    \nonumber \\ 
    &
    \quad
    \simeq \left(\frac{2 i \epsilon}{n(t-l)} \right)^{2nh_{\mathcal{O}}}\left(\frac{2 i \epsilon}{n(t-l)}\left( -\frac{t+ l}{t-l}\right)^{\frac{1}{n}} \right)^{2n \bar{h}_{\mathcal{O}}}\mathcal{F}(0, z_i^{(m)})\bar{\mathcal{F}}(0,\bar{z}_i^{(m)})
\end{align}
where $\mathcal{F}$ is the $2n$-point Virasoro conformal block. For convenience, we take the operators to be light in the $c\rightarrow \infty$ limit, in which case Wick's theorem applies
\begin{align}
    \mathcal{F}(0, z_i^{(m)})\bar{\mathcal{F}}(0,\bar{z}_i^{(m)}) = \prod_{i=1}^n\langle \mathcal{O}(z_1^{(i)}\bar{z}_1^{(i)})\mathcal{O}(z_2^{(i)}\bar{z}_2^{(i)})\rangle_{\Sigma_1},
\end{align}
so the R\'enyi entropy is 
\begin{align}
    \Delta S_n = \frac{2n h_{\mathcal{O}}}{n-1}\log \left( \frac{n(t-l)\sin\left( \frac{\pi}{n}\right)}{\epsilon}  \right).
\end{align}

\paragraph*{Twist field approach}
In the twist field approach that we have been using throughout this paper, the R\'enyi entropy is computed by
\begin{align}
    \Delta S_n  = \frac{1}{1-n} \log \frac{\langle \mathcal{O}^{\otimes n} \mathcal{O}^{\otimes n} \sigma_n \bar{\sigma}_n\rangle}{\langle\mathcal{O}^{\otimes n} \mathcal{O}^{\otimes n} \rangle \langle\sigma_n \bar{\sigma}_n\rangle},
\end{align}
where $h_{\mathcal{O}^{\otimes n}} = n h_{\mathcal{O}}$.
Using the standard conformal mapping to fix the positions of three of the operators, we have
\begin{align}
    \frac{\langle \mathcal{O}^{\otimes n} \mathcal{O}^{\otimes n} \sigma_n \bar{\sigma}_n\rangle}{\langle\mathcal{O}^{\otimes n} \mathcal{O}^{\otimes n} \rangle \langle\sigma_n \bar{\sigma}_n\rangle} = |z|^{2n\Delta_{\mathcal{O}}} \langle \sigma_n (\infty ) \bar{\sigma}_n (1) \mathcal{O}^{\otimes n}(z, \bar{z}) \mathcal{O}^{\otimes n}(0) \rangle
\end{align}
with cross-ratios
\begin{align}
    z = \frac{2 i \epsilon}{l-t+ i \epsilon}, \quad \bar{z} = -\frac{2 i \epsilon}{l+t- i \epsilon}.
\end{align}
First, we consider the entropy at times $t<l$, in which case, the four-point function can be evaluated as the vacuum ${Vir^n/\mathbb{Z}_n}$ conformal block 
\begin{align}
    \langle \sigma_n (\infty ) \bar{\sigma}_n (1) \mathcal{O}^{\otimes n}(z, \bar{z}) \mathcal{O}^{\otimes n}(0) \rangle  = \mathcal{F}^{Vir/\mathbb{Z}_n}_0(z)\bar{\mathcal{F}}^{Vir/\mathbb{Z}_n}_0(\bar{z}).
\end{align}
This is where we make our crucial assumption that we can instead evaluate the $Vir$ conformal blocks. 

Because the primary operators are light and the twist fields are heavy (for $n > 2$), we can evaluate the $Vir$ conformal blocks using techniques from Ref.~\cite{Fitzpatrick2015}
\begin{align}
    \mathcal{F}_0(z) = \left( \frac{(1-z)^{\frac{\alpha - 1}{2}}\alpha }{1 - (1-z)^{\alpha}}\right)^{n\Delta_{\mathcal{O}}} \rightarrow z^{-n\Delta_{\mathcal{O}}}, \quad \alpha = \sqrt{1 - \frac{24 h_n}{c}}.
\end{align}
This leads to 
\begin{align}
    \Delta S_n = 0
\end{align}
as is expected due to causality. Now, we consider times $t > l$ and the holomorphic conformal block picks up a monodromy $(1-z) \rightarrow e^{-2\pi i}(1-z)$ causing a nontrivial answer
\begin{align}
    \mathcal{F}_0(z) \rightarrow \left( \frac{e^{-\pi i (\alpha -1)}(1-z)^{\frac{\alpha - 1}{2}}\alpha }{1 - e^{-2\pi i \alpha }(1-z)^{\alpha}}\right)^{n\Delta_{\mathcal{O}}} \rightarrow \left(\frac{i\alpha}{2\sin(\pi \alpha)} \right)^{\Delta_{\mathcal{O}}}.
\end{align}
This gives a R\'enyi entropy of 
\begin{align}
    \Delta S_n = \frac{2nh_{\mathcal{O}}}{n-1}\log \left(\frac{(t-l ) \sin(\pi \alpha )}{\epsilon \alpha} \right).
    \label{renyi_uniform}
\end{align}
Except for the additive constant, this is precisely the same as the result from the uniformization map. One may argue that the higher R\'enyi entropies do not tell us about the von Neumann limit or the negativity because \eqref{renyi_uniform} diverges at $n=1$ (not analytic) and there can be a \textit{replica transition} \cite{2018JHEP...01..115K, 2019arXiv190502191K}. However, it has recently been shown that $O(1/c)$ corrections save the day and provide a finite von Neumann limit \cite{2020arXiv200415010A}. This gives us further confidence that the essential equivalence between the $Vir$ and $Vir^n/\mathbb{Z}_n$ blocks in the R\'enyi entropy computations of this section can carry over to the von Neumann limit and negativity.


\providecommand{\href}[2]{#2}\begingroup\raggedright\endgroup

\end{document}